\documentclass[onecolumn,twospace]{aastex62}

\usepackage{graphicx}	
\usepackage{subfigure}  
\usepackage{amsmath}	
\usepackage{amssymb}    
\graphicspath{{./}{figures/}}

\accepted{\today}
\submitjournal{ApJ}

\shorttitle{A global solution to a disk with outflows}
\shortauthors{Feng et al.}

\begin{document}

\title{A global solution to a slim accretion disk with radiation-driven outflows} 

\correspondingauthor{Xinwu Cao}

\author{Junjie Feng}
\affiliation{Key Laboratory for Research in Galaxies and Cosmology, Shanghai Astronomical Observatory, Chinese Academy of Sciences, 80 Nandan Road, Shanghai, 200030, China}
\affiliation{University of Chinese Academy of Sciences, 19A Yuquan Road, 100049,
	Beijing, China}


\author{Xinwu Cao}
\affiliation{Zhejiang Institute of Modern Physics, Department of Physics, Zhejiang University,
No.38 Zheda Road, Hangzhou 310027, China, Email: cxw@shao.ac.cn}
\affiliation{SHAO-XMU Joint Center for Astrophysics, Shanghai Astronomical Observatory, Chinese Academy of Sciences, 80 Nandan Road, Shanghai 200030}
\affiliation{University of Chinese Academy of Sciences, 19A Yuquan Road, 100049,
	Beijing, China}


\author{Wei-Min Gu}
\affiliation{Department of Astronomy, Xiamen University, Xiamen, Fujian 361005, China}

\author{Ren-Yi Ma}
\affiliation{Department of Astronomy, Xiamen University, Xiamen, Fujian 361005, China}






\begin{abstract}
The thickness of a slim disk is determined by the balance between the radiation force and the vertical component of the gravity of the black hole (BH). It was found that vertical gravity increases with the disk height, and it will decrease with the disk height if the disk thickness is above a critical value, which implies that gas at the disk surface may be driven into outflows by radiation force when the disk thickness surpasses the critical value. In this work, we derive a global solution to a slim disk with radiation-driven outflows.  We find that the outflows are driven from the disk surface if the mass accretion rate $\dot{m}\ga 1.78-1.91$  ($\dot{m}=\dot{M}/\dot{M}_{\rm Edd}$, and $\dot{M}_{\rm Edd}=L_{\rm Edd}/0.1c^2$) depending on BH mass, while the outflows are suppressed when the mass accretion rate is lower than this critical value. The mass accretion rate decreases with decreasing radius in the disk with outflows, and the rate of the gas swallowed by the BH is always limited to $\dot{m}_{\rm in}\sim 1.78-1.91$ even if the mass accretion rate at the outer edge of the disk is very high. This may set constraints on the massive BH growth through accretion in the early Universe. Due to the presence of outflows, there is an upper limit on the radiation flux of the disk, which leads to saturation of the continuum spectra of the disk with outflows at the high energy band. This may be tested by the observations of quasars or/and BH X-ray binaries.
\end{abstract}



\section{Introduction} \label{sec:intro}

The accretion onto black holes (BHs) is the dominant energy source of different kinds of astrophysical objects, such as, active galactic nuclei (AGNs), black hole X-ray binaries (BHXBs) and tidal disruption events (TDEs). The thin accretion disk model is widely used for the sources with moderate luminosity, $L_{\rm bol}\ll L_{\rm Edd}$ ($L_{\rm bol}$ is the bolometric luminosity, and $L_{\rm Edd}$ is the Eddington luminosity) \citep*[][]{1973A&A....24..337S,1989MNRAS.238..897L}. In the thin disk model, the vertical component of the gravity of the BH is assumed to be in equilibrium with vertical pressure gradient of the disk, where the vertical component of the gravity, $GMz/R^3$, is a good approximation because of its thickness $H/R\ll 1$, as the disk luminosity is sub-Eddington \citep*[][]{1973A&A....24..337S}. Almost all the released gravitational energy is radiated out locally from a thin accretion disk. The radiation pressure increases with mass accretion rate, and the disk expands vertically. {The disk thickness can be comparable with radius if the accretion rate is high}, and therefore the thin disk approximation for vertical gravity is invalid.
The disk is slim or even geometrically thick, and photons are trapped at some specific radii and a substantial fraction of the gravitational energy is advected into the BH \citep{1980AcA....30....1J,1980A&A....88...23P,1988ApJ...332..646A}. The slim disk model can describe BHs
accreting at super-Eddington accretion rates ($\dot M_{\rm{Edd}} = L_{\rm{Edd}} / \eta c^2$, $L_{\rm{Edd}} = 4\pi cGM_{\rm{BH}}/\kappa_{\rm{es}}$). The slim disk model has been extensively investigated in many previous works \citep*[e.g.,][]{1999PASJ...51..725W,2001PASJ...53..915W,2007ApJ...660..541G,2009ApJ...693..670J,2011A&A...527A..17S,2011MNRAS.413.1623D}, which can successfully reproduce the main observational features of super-Eddington accreting BHs  \citep*[e.g.,][]{2000PASJ...52..499M,2001ApJ...549L..77W,2003A&A...398..927W,2009ApJ...698.1515D,2009MNRAS.398.1905W,2014ApJ...797...65W,2018PASJ...70..108K}.

There is observational evidence of outflows from super-Eddington accretion disks {\citep*[e.g.,][]{2009MNRAS.397.1836G,2011MNRAS.417..464M,2015ApJ...806...22D}}, which is also confirmed by numerical simulations \citep*[e.g.,][]{2014ApJ...780...79Y,2014ApJ...796..106J,2018ApJ...867..100Y}. The theoretical works also suggest that outflows may be driven from the surface of a slim disk \citep{2006ApJ...652..518M,2007ApJ...660..541G,2009ApJ...693..670J,2012ApJ...753..118G,2015MNRAS.448.3514C}. They pointed out that the widely used approximation for vertical component of gravity $GMz/R^3$ is only valid for the thin disk with $H/R\ll 1$, and a more accurate expression of vertical gravity should be adopted for a slim disk. \citet{2006ApJ...652..518M} found that the radiation force would be inevitably dominant over the vertical gravitational force at disk surface if the relative half-thickness of the disk is $H/R>1/\sqrt{2}$. This implies that the gas at the disk surface will be blown away by the radiation force. \citet{2015MNRAS.448.3514C} investigated the structure of such a super-Eddington accretion disk with outflows accelerated by the radiation force. Their analytic model of an accretion disk with outflows is developed based on two assumptions, i.e., Keplerian motion of the gas in the disk and neglecting of the radial energy advection in the disk. A Newtonian gravitational potential is adopted in their model calculations, which is similar to the standard thin accretion disk, and therefore the model results may not be able to accurately describe the inner region of the disk very close to the BH. As most of the gravitational energy of the gas is released in the inner region of the disk, a global transonic structure of such a slim disk surrounding a BH is crucial in modeling the observations.


In this work, we derive a global solution to a slim accretion disk with outflows accelerated by radiation force of the disk. The calculations can be applied to model the accretion disks surrounding either massive BHs or stellar mass BHs. We describe our model calculations and results in Sections \ref{sec:model} and \ref{sec:results}.
The last section contains the discussion of the results.

\section{Model} \label{sec:model}

A set of equations describing a slim disk with outflows in cylindrical coordinates are summarized in the previous works \citep[see][]{2008bhad.book.....K,2007ApJ...660..541G}. We consider a steady axisymmetric accretion disk in this work, which is described by the momentum equations, energy equation, continuity equation, and the equation of state.

The $r$-component of the momentum equation is
\begin{equation}
    \frac{\partial}{r\partial{r}}(r\rho v^2_r)+\frac{\partial}{\partial{z}}(\rho v_r v_z)-\frac{\rho v^2_{\phi}}{r}=-\frac{\partial{p}}{\partial{r}}-\rho \frac{\partial{\psi}}{\partial{r}}.
	\label{eq:r-momentum equation1}
\end{equation}
As done in many previous works, we adopt the pseudo-Newtonian potential,
\begin{equation}
    \psi=-\frac{GM}{\sqrt{r^2+z^2}-r_{\rm{s}}},
	\label{eq:pwpotential}
\end{equation}
to simulate the gravity of the BH, where $M$ is the BH mass, and $r_{\rm{s}}\equiv2GM/c^2$ is the Schwarzschild radius \citep*[][]{1980A&A....88...23P}.

The continuity equation is
\begin{equation}
    \frac{\partial}{r\partial{r}}(\rho rv_r)+\frac{\partial}{\partial{z}}(\rho v_z)=0,
	\label{eq:continuity1}
\end{equation}

We integrate Equation (\ref{eq:r-momentum equation1}) over $z$ and obtain
\begin{equation}
    v_r\frac{dv_r}{dr}+\frac{1}{\Sigma}\frac{d\Pi}{dr}+r(\Omega^2_{\rm K}-\Omega^2)-\frac{1}{r}\frac{\Pi}{\Sigma}(\frac{3}{2}+\frac{r_{\rm s}}{r-r_{\rm s}})=0,
	\label{eq:r-momentum equation2}
\end{equation}
where Equation (\ref{eq:continuity1}) is used, $\Sigma$ is the surface density, $\Pi$ is the vertically integrated pressure, $\Omega$ is the angular velocity of the gas in the disk, and $\Omega_{\rm K}$ is the Keplerian angular velocity. The $\phi$-component of the momentum equation is
\begin{equation}
    \frac{\partial}{r\partial{r}}(r\rho v_rv_{\phi})+\frac{\partial}{\partial{z}}(\rho v_{\phi} v_z)+\frac{\rho v_r v_{\phi}}{r}=\frac{\partial}{r^2\partial{r}}(r^2t_{r\phi}),
	\label{eq:phi-momentum equation1}
\end{equation}
where the $\alpha$-viscosity $t_{r\phi}=-\alpha p$ is adopted. Integrating Equation (\ref{eq:phi-momentum equation1}) over $z$-direction, we have
\begin{equation}
    \frac{\partial}{r\partial{r}}(r\Sigma v_rv_{\phi})+2\rho v_{\phi} v_H+\frac{\Sigma v_r v_{\phi}}{r}=\frac{\partial}{r^2\partial{r}}(-r^2\alpha \Pi),
	\label{eq:phi-momentum equation2}
\end{equation}
where the velocity of outflow at the disk surface $v_{H}=v_z(H)$, as the outflows driven by the radiation force are considered in our model.

Integrating Equation (\ref{eq:continuity1}) vertically, it becomes
\begin{equation}
    2\rho v_{\rm H}=\frac{1}{r}\frac{\partial(r\Sigma v_r)}{\partial r}.
	\label{eq:rhovz}
\end{equation}
The term $2\rho v_{\rm H}$ in this equation represents the mass loss rate in outflows from unit area of the disk surface. We further integrate Equation (\ref{eq:rhovz}) over $r$-direction, it becomes
\begin{equation}
r\Sigma v_r+\int 2\rho v_{\rm H}rdr=C, \label{eq:continuity3}
\end{equation}
where the integral constant $C$ can be determined with suitable boundary conditions. As the mass loss rate in the outflows from unit area of the disk surface $\dot{m}_{\rm w}\equiv2\rho v_{\rm H}$, we have
\begin{equation}
    \dot{M}(r)=\dot{M}_{\rm{out}}-\dot{M}_{\rm{w}}(r)=-2\pi r\Sigma v_r,
	\label{eq:continuity4}
\end{equation}
so the mass accretion rate in the disk at radius $r$ is
\begin{equation}
\dot{M}(r)=-2\pi r\Sigma v_r,
\end{equation}
and the total mass loss rate in the outflows driven from the disk region between $r$ and $r_{\rm out}$ is
\begin{equation}
\dot{M}_{\rm w}(r)=\int\limits_{r}^{r_{\rm out}}2\pi r\dot{m}_{\rm w}dr.
\end{equation}

Substituting Equation (\ref{eq:rhovz}) into Equation (\ref{eq:phi-momentum equation2}), we have
\begin{equation}
    \frac{\partial}{r\partial{r}}(r\Sigma v_rv_{\phi})+ \frac{v_{\phi}}{r}\frac{\partial(r\Sigma v_r)}{\partial r}+\frac{\Sigma v_r v_{\phi}}{r}=\frac{\partial}{r^2\partial{r}}(-r^2\alpha \Pi).
	\label{eq:phi-momentum equation3}
\end{equation}
We integrate the angular equation (\ref{eq:phi-momentum equation3}) over $r$, which leads to
\begin{equation}
    \dot {M}(r) \Omega r^2 -\dot{M}_{\rm{in}}j_{\rm {in}} - \int\limits^{r}_{r_{\rm{in}}} \Omega r^2 d\dot{M} = 2\pi\alpha r^2 \Pi,
	\label{eq:phi-momentum equation4}
\end{equation}
respectively, where Equation (\ref{eq:continuity4}) is used, $\int^{r}\limits_{r_{\rm{in}}} \Omega r^2d\dot{M}$ is the angular momentum removed by the outflows from the corresponding disk region, $\dot{M}_{\rm{in}}$ is the mass accretion rate swallowed by black hole, and $\dot{M}_{\rm{in}}j_{\rm in} $ is rate of angular momentum of the accretion gas flowing into the black hole ($j_{\rm in}$ is the specific angular momentum of the accreting gas swallowed by the BH).

The vertical integrated equation of state is
\begin{equation}
    \Pi = \frac{8}{9}\frac{\Sigma k_B}{\mu m_p}T_0 +\frac{256}{945}aHT^4_0,
	\label{eq:equation of state}
\end{equation}
where $\mu$ is the mean molecular weight ($\mu=0.62$ is used in this work), and $T_0$ is the temperature at the mid-plane of the disk \citep[see Equations 7.43, 7.44 and 7.45 in][]  {2008bhad.book.....K}. The isothermal sound speed is $c^2_{\rm{s}} = \Pi/\Sigma$.


The energy equation of the disk is
\begin{equation}
    Q^+_{\rm{vis}}=Q^-_{\rm{adv}}+Q^-_{\rm{rad}},
	\label{eq:equation of energy}
\end{equation}
where $Q^+_{\rm{vis}}$ is the gravitational power released viscously in unit area of the disk, $Q^-_{\rm{rad}}$ is the cooling rate due to radiation, and $Q^-_{\rm{adv}}$ is the advection term. They are given by
\begin{equation}
    Q^+_{\rm{vis}} = \frac{3}{2}\alpha \Pi \Omega,
	\label{eq:viscosity energy}
\end{equation}
\begin{equation}
    Q^-_{\rm{adv}} = \frac{1}{r}\frac{d}{dr}(4rv_r\Pi)-v_r \frac{d\Pi}{dr}+\Pi v_r \frac{d\ln H}{dr},
	\label{eq:advective energy}
\end{equation}
and
\begin{equation}
    Q^-_{\rm{rad}} = \frac{8acT^4_0}{105/32\overline{k}\Sigma},
	\label{eq:radiative energy}
\end{equation}
where the opacity is $\overline{k}=k_{\rm{es}}+k_{\rm{ff}}=0.34+6.4 \times 10^{22} \overline{\rho}\overline{T}^{-7/2} {\rm cm}^2 \ {\rm g}^{-1}$, $\overline{\rho}(=\Sigma /2H)$ and $\overline{T}(=2T_0/3)$ are the vertically averaged density and temperature respectively \citep*[see][for the details]{2008bhad.book.....K}.


The vertical structure of an accretion disk has been investigated in detail by \citet{1996astro.ph.11101A}. A general accurate expression for the vertical hydrodynamical equilibrium valid both for a thin and a slim disk has been derived in their work. The hydrostatic balance in the vertical direction of the accretion disk is described by
\begin{equation}
    \frac{1}{\rho}\frac{\partial{p}}{\partial{z}}+\frac{\partial{\psi}}{\partial{z}}+v_r\frac{\partial{v_z}}{\partial{r}}+v_z\frac{\partial{v_z}}{\partial{z}}=0,
	\label{eq:hydrostatic equbriliumO}
\end{equation}
where $p$ is the pressure, $\rho$ is the density of the gas, $\psi$ is the gravitational potential  \citep*[][]{1996astro.ph.11101A}. As we focus on how the general properties of the disk structure is affected by the radiation driven outflows, we simplify the expression of the hydrostatic balance in the vertical direction as
\begin{equation}
    \frac{1}{\rho}\frac{\partial{p}}{\partial{z}}+\frac{\partial{\psi}}{\partial{z}}=0,
	\label{eq:hydrostatic equbrilium}
\end{equation}
 where we assume the terms of $\partial v_z/\partial r$ and $\partial v_z/\partial z$ to be negligible. It has been pointed out that the widely used approximation for vertical component of gravity $GMz/R^3$ is only valid for the thin disk with $H/R\ll 1$, and a more accurate expression of vertical gravity should be adopted for a slim disk \citep{2006ApJ...652..518M,2007ApJ...660..541G,2009ApJ...693..670J,2012ApJ...753..118G,2015MNRAS.448.3514C}.
As discussed in \citet{2015MNRAS.448.3514C}, there are upper-limits on the radiation flux $f_{\rm{rad}}$ and half-thickness $H$ for a slim disk, above which, the radiation force will overwhelm the vertical gravity. The maximal flux can be calculated with 
\begin{equation}
    f_{\rm{rad}}=q(H)=\frac{GM\rho H}{\sqrt{r^2+H^2}(\sqrt{r^2+H^2}-r_{\rm s})^2}.
	\label{eq:outflux}
\end{equation}
How the gas is blown away from the disk surface by the radiation force is still quite unclear, and therefore it is assumed that the gas at the disk surface will be accelerated into outflows by the radiation force when the radiation force overwhelms the vertical gravity. Then the mass accretion rate in the disk decreases due to outflows. This self-adjustment mechanism leads to a maximal radiation flux/thickness of the disk \citep*[see the detailed discussion in][]{2015MNRAS.448.3514C}. 
Thus, we assume that the outflows are triggered in the disk where the condition,
\begin{equation}
    Q^-_{\rm{rad}}\ge2f^{\rm{max}}_{\rm{rad}},
    \label{eq:outflow condition}
\end{equation}
is satisfied, where $f^{\rm{max}}_{\rm{rad}}$ is the maximal radiation flux and then the mass accretion rate is self-adjusted by the outflows to maintain
\begin{equation}
    Q^-_{\rm{rad}}\equiv 2f^{\rm{max}}_{\rm{rad}}.
\label{q_rad_f_rad}
\end{equation}
This relation is adopted in the energy equation (\ref{eq:equation of energy}) for the region in the disk where outflows are driven. Otherwise, the outflows are suppressed, and the disk is the same as the conventional slim disk without outflows, which is described by a set of equations for a  slim disk \citep*[see Chapter 7 in][]{2008bhad.book.....K}, i.e., the continuity equation,
\begin{equation}
    \dot{M}=-2\pi r\Sigma v_r,
	\label{eq:2continuity}
\end{equation}
where $\dot{M}$ remains constant radially,
and the angular momentum equation,
\begin{equation}
    v_r (\Omega r^2-j_{\rm in})=-\alpha r c^2_{\rm s}.
	\label{eq:2phi-momentum equation}
\end{equation}
{Except these three equations (\ref{q_rad_f_rad}, \ref{eq:2continuity} and \ref{eq:2phi-momentum equation}), the radial-component of the momentum equation, the equation of state, and energy equation are the same as the disk with outflows (see Equations \ref{eq:r-momentum equation2}, \ref{eq:equation of state}, \ref{eq:equation of energy}, \ref{eq:viscosity energy}, \ref{eq:advective energy}, and \ref{eq:radiative energy}). }

Now, the global structure of a slim disk with outflows is available by integrating two differential equations (Equations \ref{eq:r-momentum equation2} and \ref{eq:equation of energy}) inwards from the outer radius of the disk with suitable boundary conditions. We define the dimensionless mass accretion rate in the disk as $\dot{m}=\dot{M}/\dot{M}_{\rm Edd}$, where $\dot{M}_{\rm Edd}=L_{\rm Edd}/\eta c^2$ ($\eta=0.1$ is adopted).



With the derived global structure of the disk with outflows, the effective temperature $T_{\rm{eff}}$ of the disk as a function of radius is available,
\begin{equation}
T_{\rm eff}(r)= \left[{\frac {Q^-_{\rm rad}(r)}{2\sigma}}\right]^{1/4}.\label{t_eff}
\end{equation}
The continuum spectrum of the disk can be calculated with
\begin{equation}
    L_{\nu}=2\int \frac{h\nu^3}{c^2}\frac{2\pi rdr}{e^{h\nu/k_{\rm{B}}T_{\rm{eff}}}-1},
	\label{eq:spectra}
\end{equation}
based on the derived global structure of the disk, where $h$ is Planck constant.

\section{Results} \label{sec:results}

The model has three parameters, i.e., the viscosity parameter $\alpha$, the outer radius of the disk $r_{\rm out}$, and the mass accretion rate $\dot{m}_{\rm out}$ at $r_{\rm out}$. In this work, we adopt a typical value of $\alpha=0.1$ in all calculations. The calculations are carried out for two cases: a super-massive black hole with $M=10^8M_{\odot}$ and a stellar mass black hole with $M=10M_{\odot}$, respectively. As the outer region of the disk is suffered from the gravitational instability \citep*[][]{1987Natur.329..810S,2003MNRAS.339..937G,2008A&A...477..419C}, the size of the disk is constrained by such instability. The previous calculations show that the size of the disks in luminous quasars is $\sim10^3-10^4r_{\rm s}$ \citep*[][]{2012ApJ...761..109Y}, while it is $\sim10^4-10^5r_{\rm s}$ for a disk surrounding a stellar mass BH \citep*[][]{1989MNRAS.238..897L}. Thus, we take $r_{\rm out}=10^3r_{\rm s}$ for the super-massive BH and $r_{\rm out}=10^4r_{\rm s}$ for a stellar mass black hole, respectively. The calculations are carried out by integrating two differential equations (Equations \ref{eq:r-momentum equation2} and \ref{eq:equation of energy}) from the the outer radius $r_{\rm out}$ inwards, and the parameter $j_{\rm in}$ is tuned till the derived global solution passes the sonic point smoothly.

\begin{figure}
\centering
\includegraphics[width=8cm,clip=]{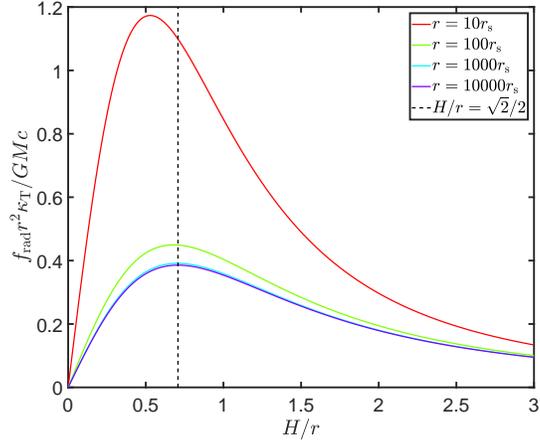}
\caption{The radiation flux varies with disk thickness at different radii. }
\label{fig:frad_h}
\end{figure}

\begin{figure*}

    \centering
    \includegraphics[width=8cm,clip=]{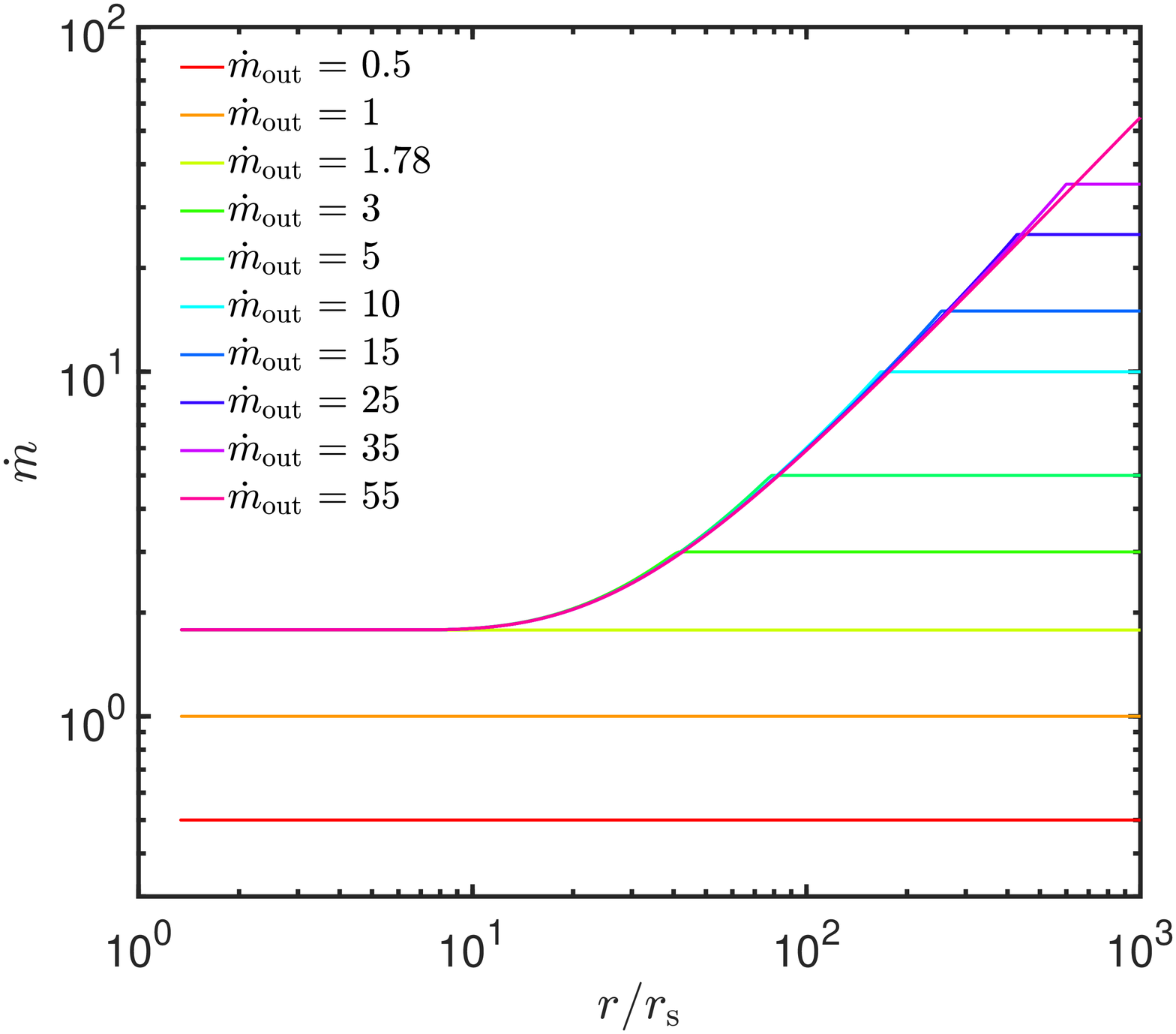}
    \includegraphics[width=8cm,clip=]{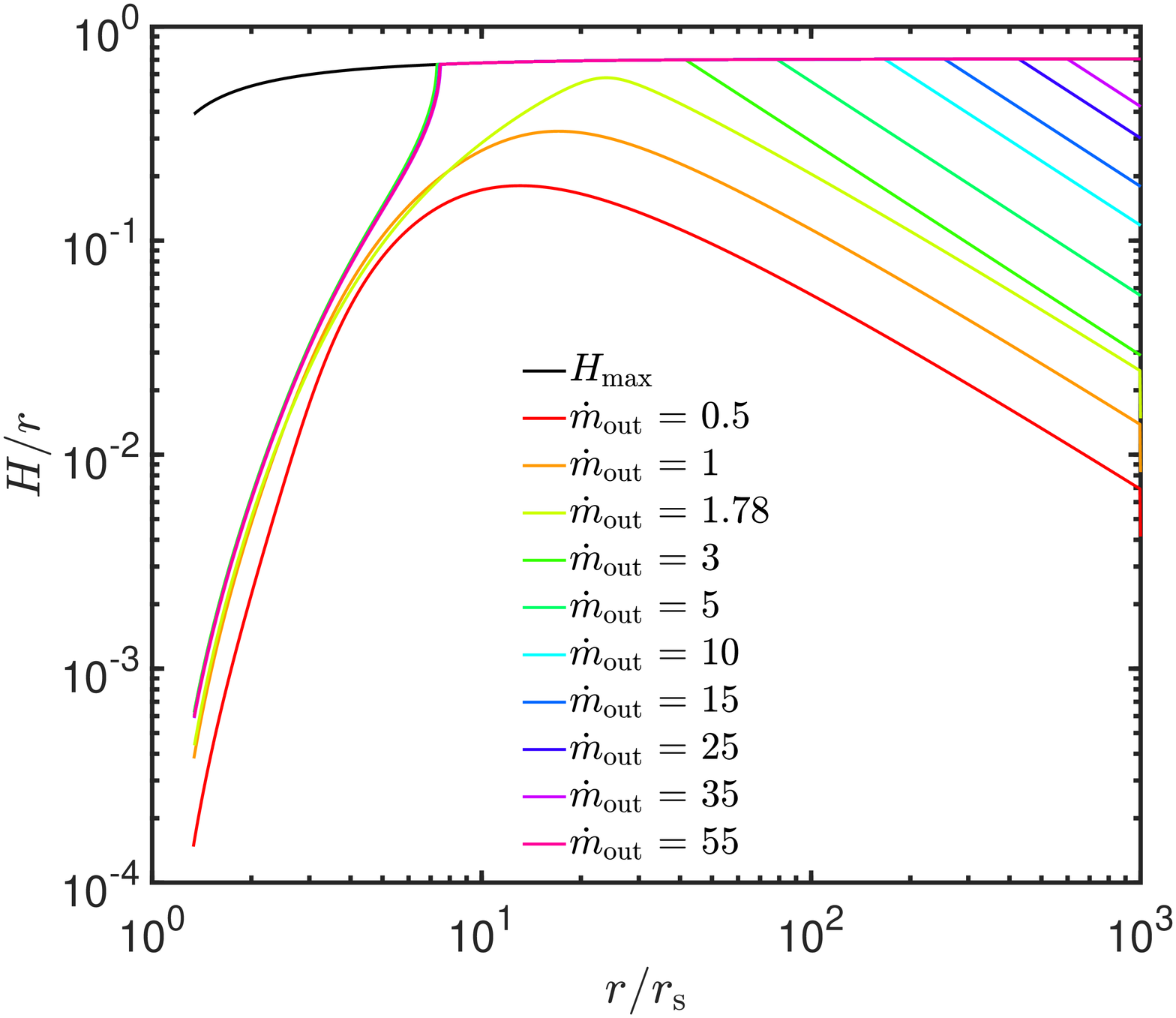}
    \caption{The mass accretion rate and the disk thickness as functions of radius for different values of the accretion rate at $r_{\rm out}$ (different color lines). The BH mass $M=10^8M_\odot$ is adopted. The black solid line in the right figure represents the maximal half-thickness of the disk calculated with Equation (\ref{eq:outflux}).}
    \label{fig:m&h1e9} 
\end{figure*}

\begin{figure*}
\centering
\includegraphics[width=8cm,clip=]{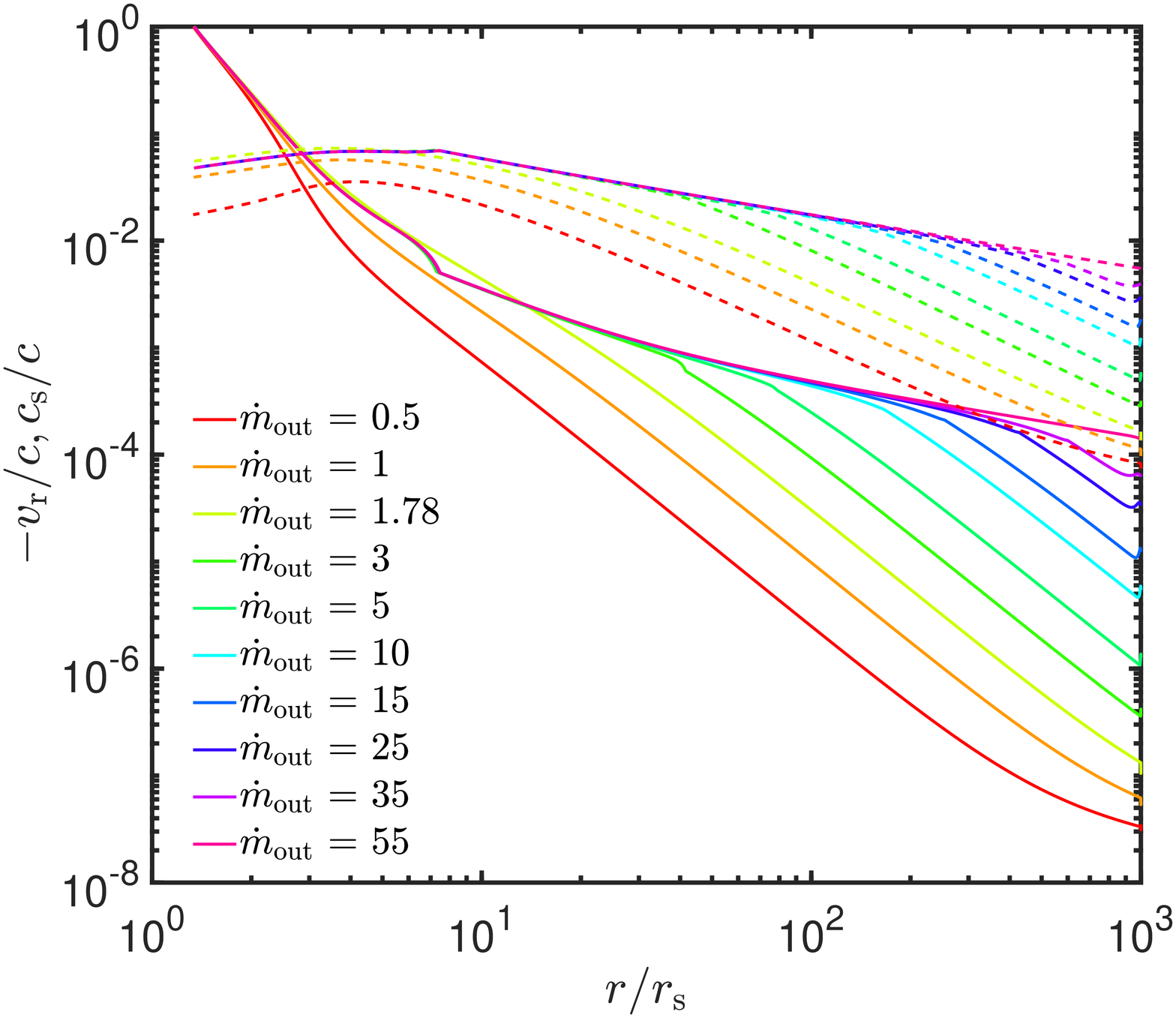}
\includegraphics[width=8cm,clip=]{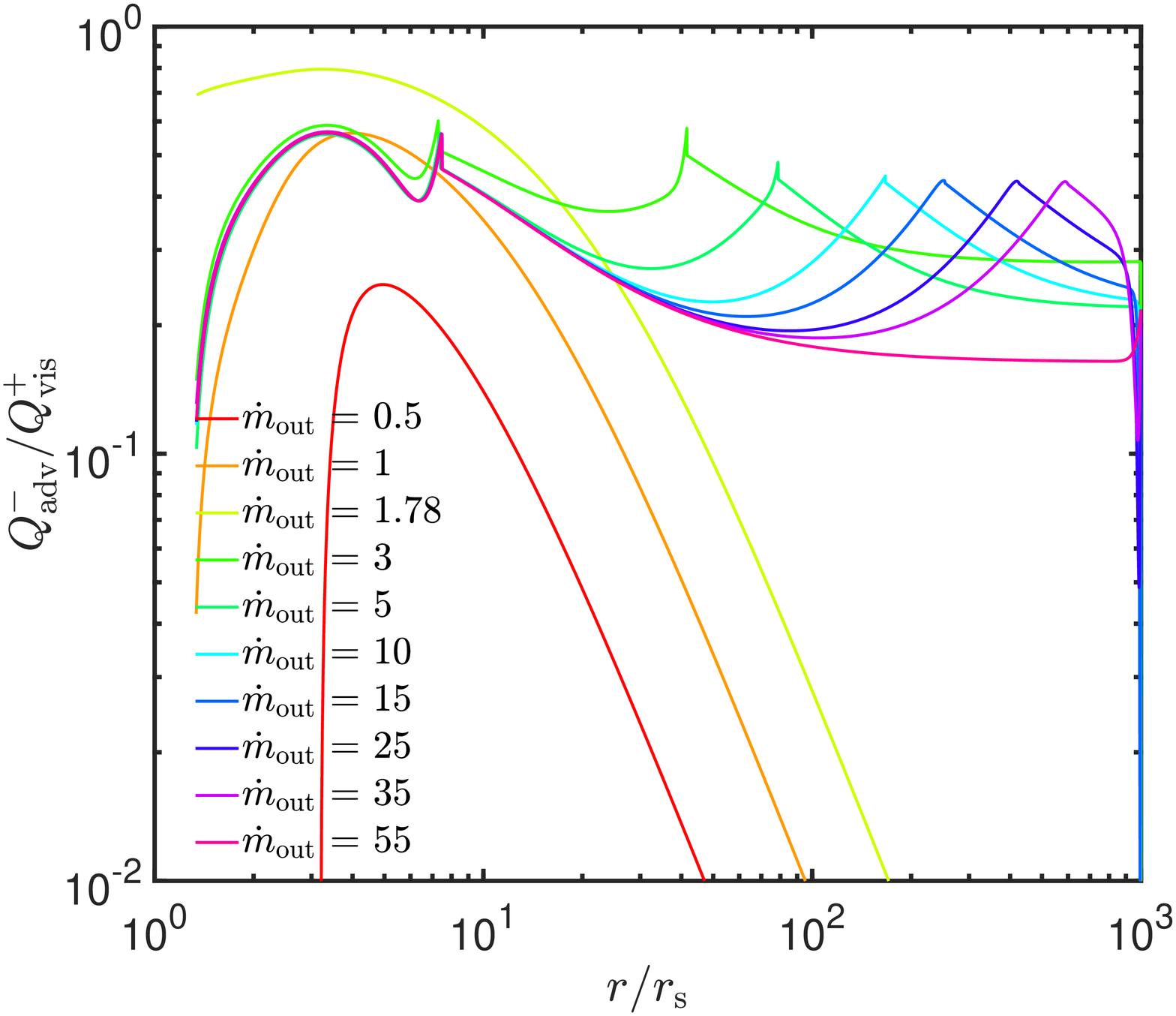}
\caption{The left figure: the same as Figure \ref{fig:m&h1e9}, but for the radial velocity and sound speed of the disk. The solid lines represent the radial velocity of the gas, while the dashed lines denote the sound speed. The right figure: the ratio of the energy advection to the viscously dissipated energy in the disk.}
\label{fig:vrcs1e9}
\end{figure*}

\begin{figure*}
	\centering
	\includegraphics[width=8cm,clip=]{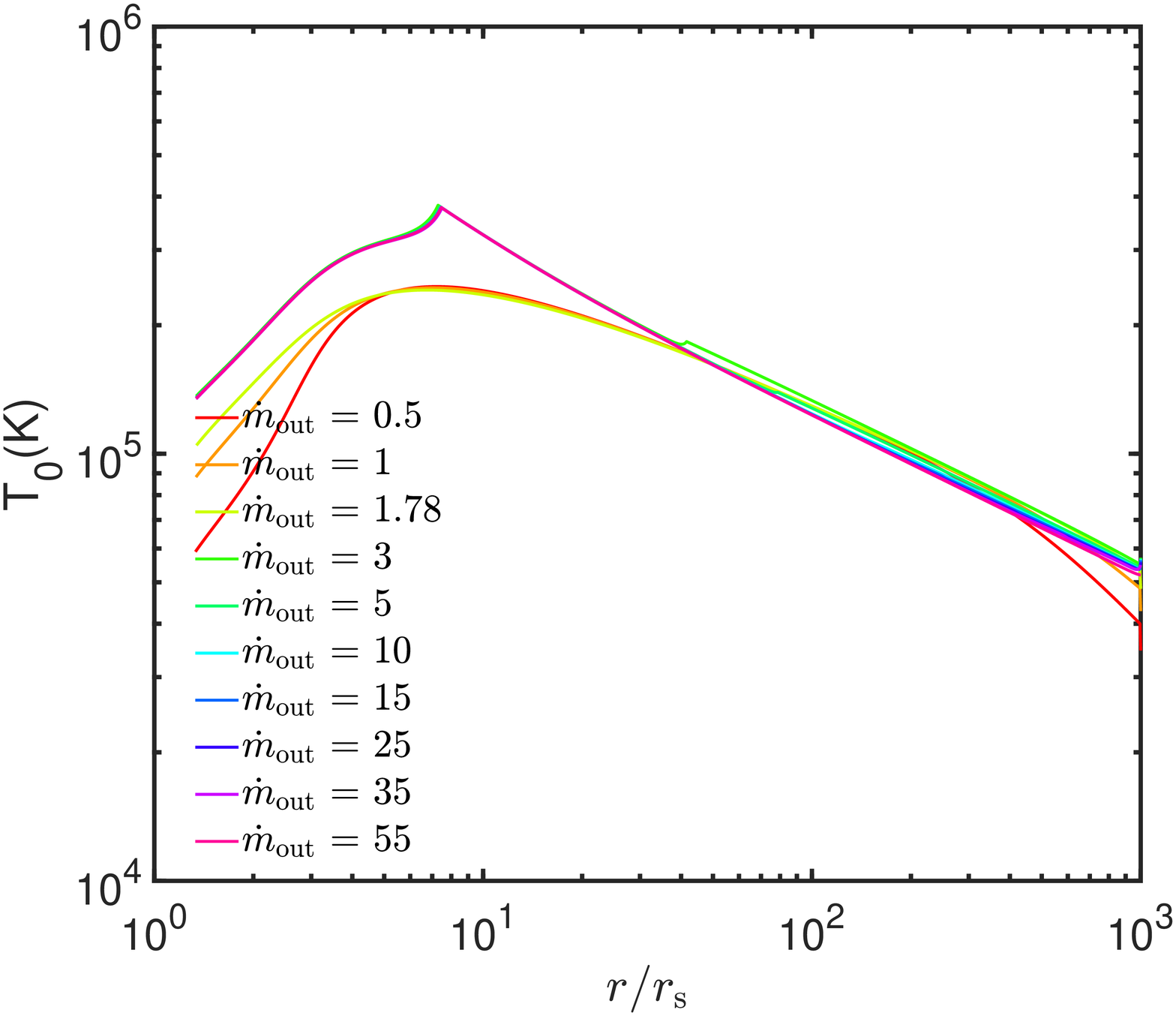}
	\includegraphics[width=8cm,clip=]{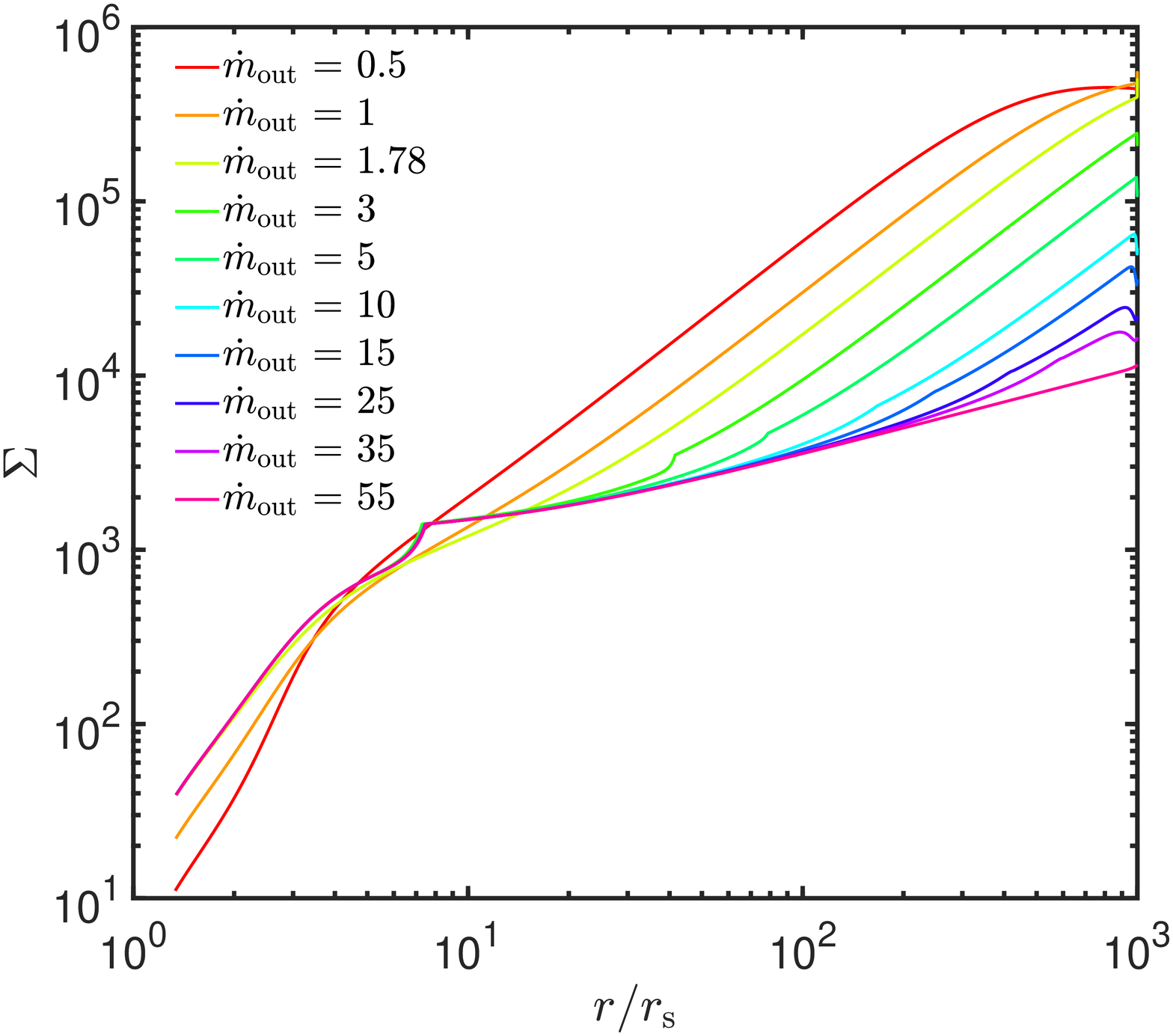}
    \includegraphics[width=8cm,clip=]{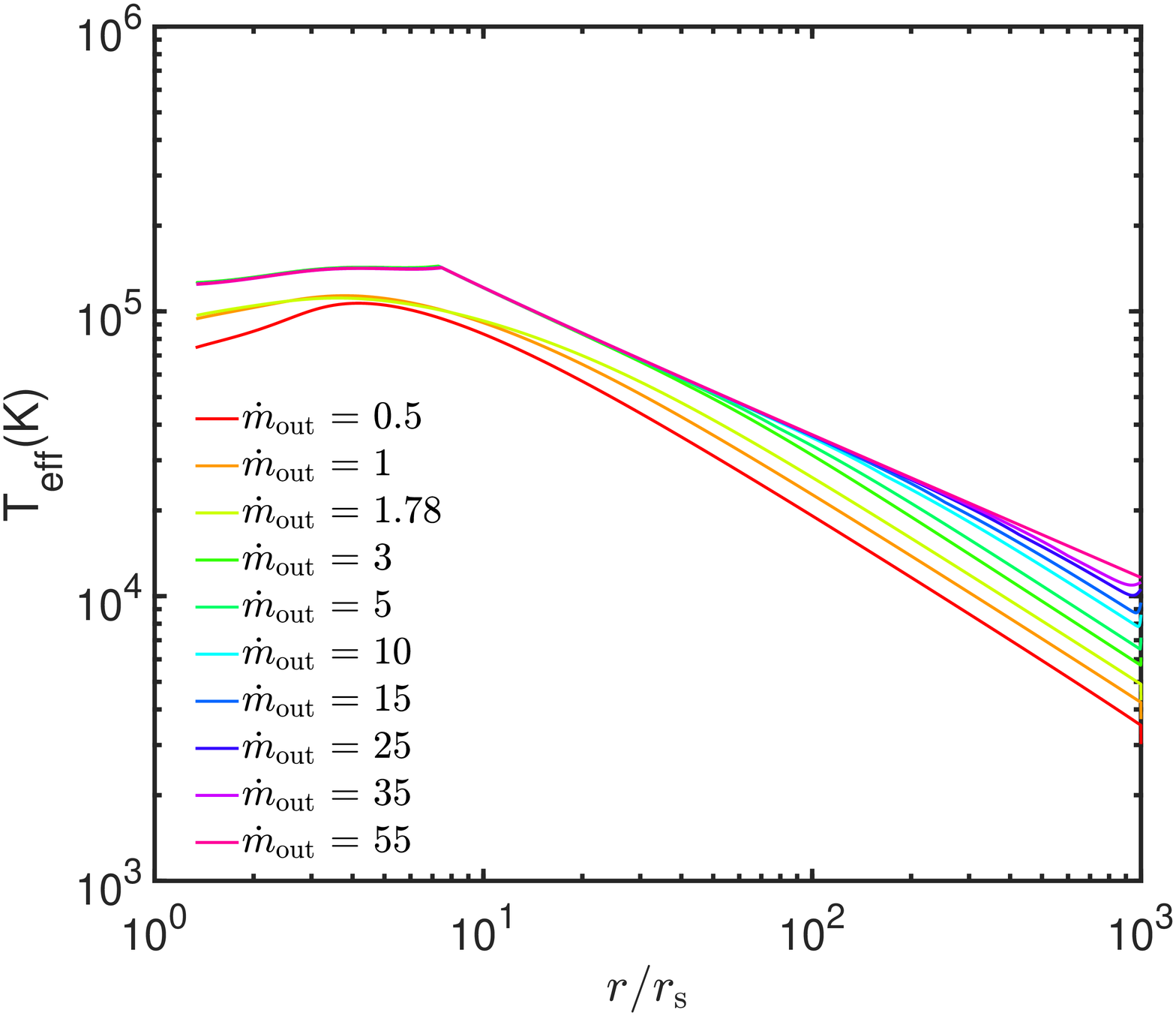}
	\includegraphics[width=8cm,clip=]{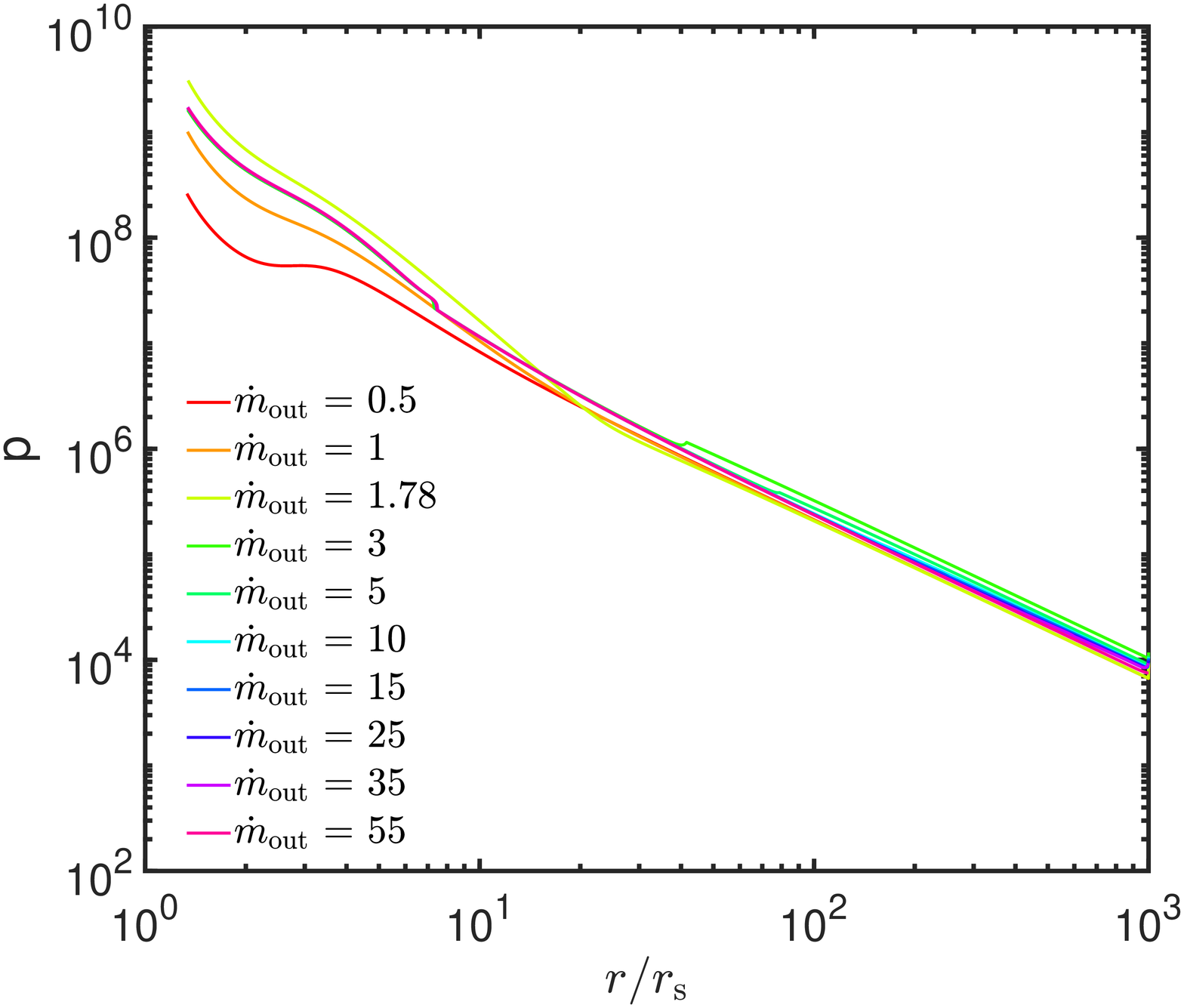}
    \caption{The same as Figure \ref{fig:m&h1e9}, but for the temperature at the disk mid-plane, the effective temperature, the surface density, and the density of the disk, respectively. }
    \label{fig:state}
\end{figure*}

In Figure \ref{fig:frad_h}, we plot the radiation flux varying with disk thickness at different radii for a vertical hydrostatic disk (see Equation \ref{eq:outflux}). There is a maximal flux at a given radius, which means that the vertical hydrostatic balance is destroyed and outflows are inevitably  driven from the disk surface if $f_{\rm rad}>f_{\rm rad}^{\rm max}$  \citep*[see][for the detailed discussion]{2015MNRAS.448.3514C}. The maximal flux corresponds to $H_{\rm max}/r\approx \sqrt{2}/2$ if $r\gg r_{\rm s}$, which reduces to the analytic result given in \citet{2015MNRAS.448.3514C}, while the maximal disk thickness decreases in the region near the BH. We plot the global structure of a slim disk with outflows for a massive black hole with different values of the mass accretion rate at the outer radius in Figures \ref{fig:m&h1e9}-\ref{fig:state}.

We find that most gas in the disk is driven into the outflows by the radiation force of the disk, if the mass accretion rate at the outer radius of the disk is sufficiently high (see Figure \ref{fig:m&h1e9}). If the mass accretion rate at the outer radius is lower than a critical value $\dot{m}_{\rm crit}\simeq 1.78$, the radiation from the disk is too weak to drive outflows from the disk surface. When the mass accretion rate $\dot{m}_{\rm out}>\dot{m}_{\rm crit}$, outflows are accelerated from a ring in the disk. The width of the ring increases with the mass accretion rate at the outer radius (see Figure \ref{fig:m&h1e9}). If $\dot{m}_{\rm out}$ is as high as 50, the ring responsible for outflows may extend to several hundred Schwarzschild radii. We note that no outflows are driven from the outer region close to the outer radius of the disk even if $\dot{m}_{\rm out}$ is very large. As the gas in the disk is driven into outflows, which makes mass accretion rate decrease with decreasing radius, and mass accretion rate onto the BH always remains around the critical mass accretion rate $\dot{m}_{\rm crit}$. {It is found that the radial velocity of the disk with outflows is higher than that of a conventional slim disk, which leads to a higher ratio of the advection to the viscous energy except in the inner region of the disk (see Figure \ref{fig:vrcs1e9}).}

\begin{figure}
\centering
\includegraphics[width=8cm,clip=]{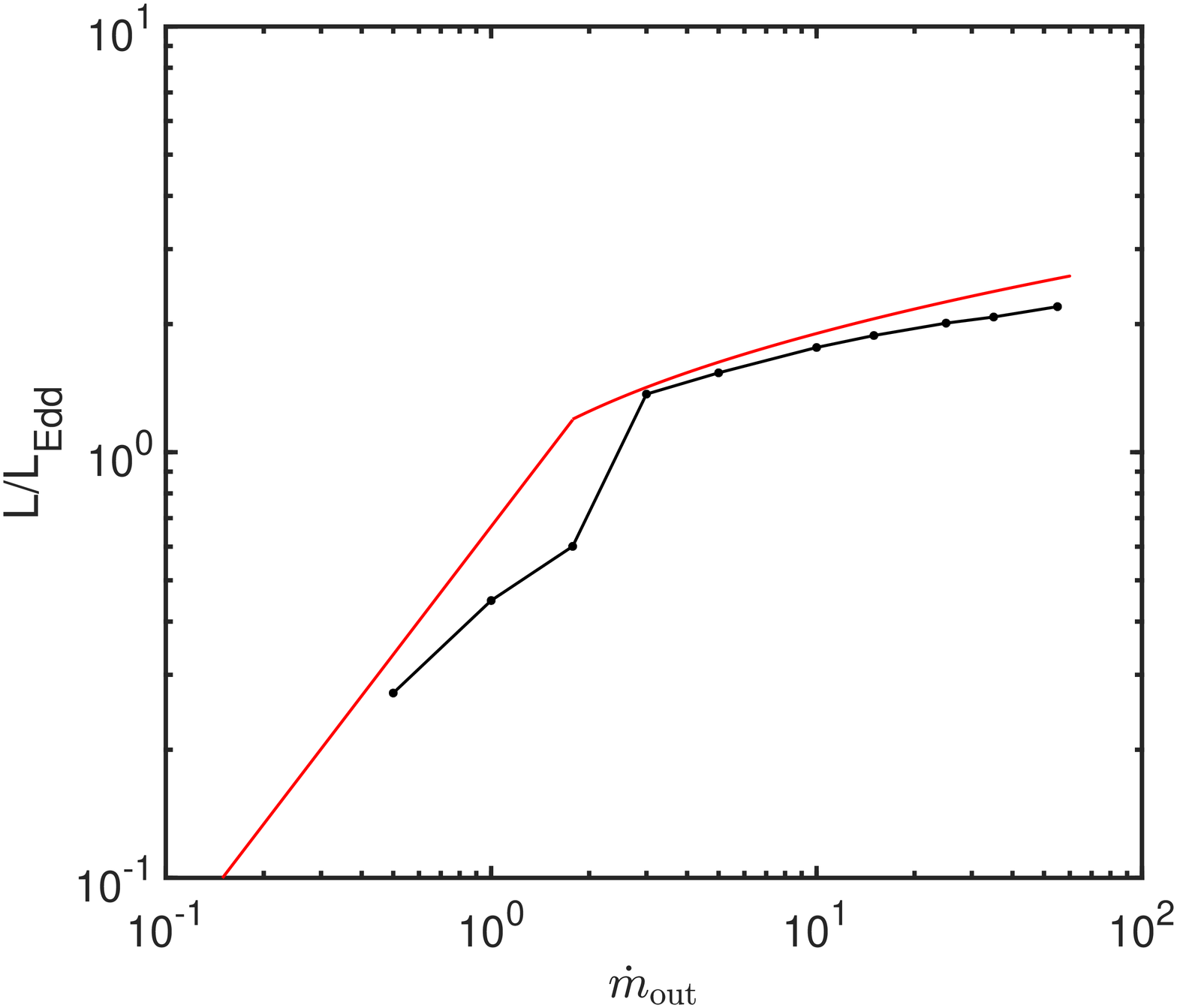}
\caption{The Eddington ratio varies with the mass accretion rate $\dot{m}_{\rm out}$ at $r_{\rm out}$. The black line represents the numerical result of the global structure of the disk, while the red line is the result calculated with the analytical approximation (24) in \citet{2015MNRAS.448.3514C}. }
\label{fig:L1e9}
\end{figure}

If the mass accretion rate at the outer radius $\dot{m}_{\rm out}<\dot{m}_{\rm crit}$, no outflows are accelerated from the disk surface, and the disk is almost the same as a normal slim disk accreting at a low rate. It is found that their temperature at the mid-plane is substantially lower than those of the disks with outflows, while their effective temperature does not deviate much from those of the disks with outflows except in the inner region of the disk (see Figure \ref{fig:state}), because the radiation in the outflow driving region is automatically regulated by the outflows, which sets an upper limit on the effective temperature. In Figure \ref{fig:L1e9}, we plot the Eddington ratio varying with the mass accretion rate $\dot{m}_{\rm out}$. The numerical result of the global structure of the disk derived in this work is qualitatively consistent with the analytical approximation (24) in \citet{2015MNRAS.448.3514C}, which is derived based on a Newtonian gravitational potential.

\begin{figure*}
    \centering

    \includegraphics[width=8cm,clip=]{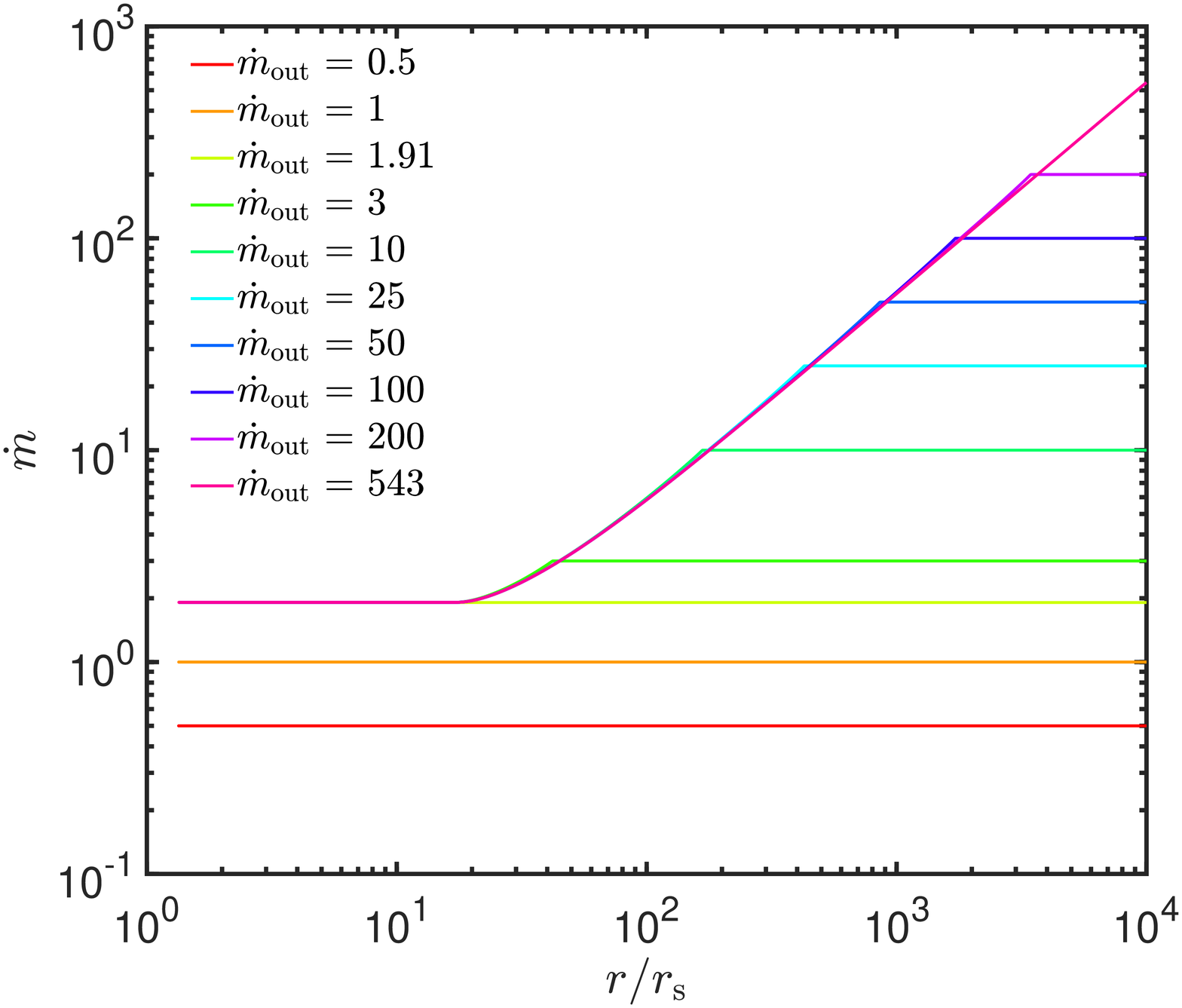}
    \includegraphics[width=8cm,clip=]{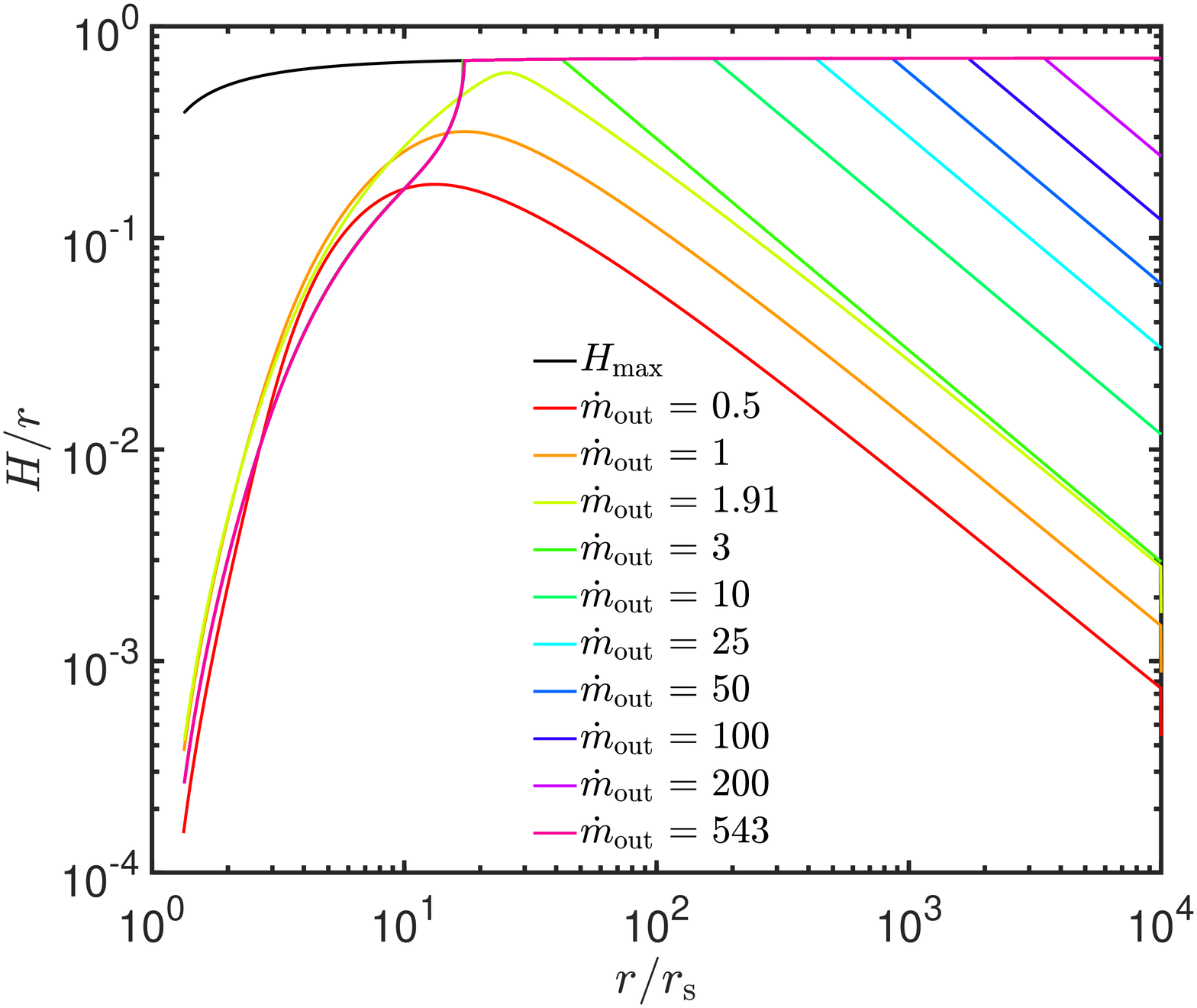}

    \caption{The mass accretion rate and the disk thickness as functions of radius for different values of the accretion rate at $r_{\rm out}$ (different color lines). The BH mass $M=10M_\odot$ is adopted. The black solid line in the right figure represents the maximal half-thickness of the disk calculated with Equation (\ref{eq:outflux}).
    }
    \label{fig:m&h10}
\end{figure*}

\begin{figure*}
\centering
\includegraphics[width=8cm,clip=]{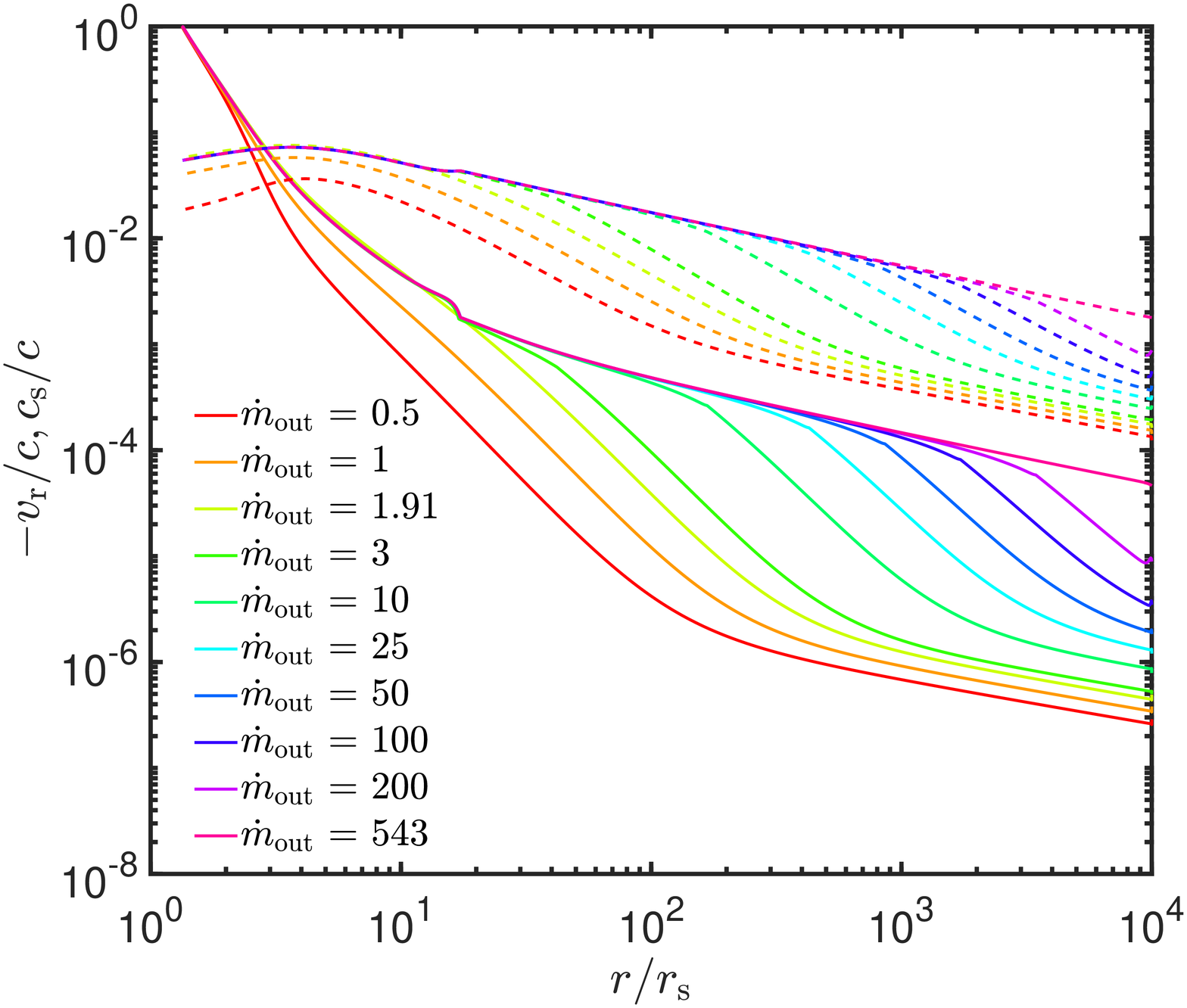}
\includegraphics[width=8cm,clip=]{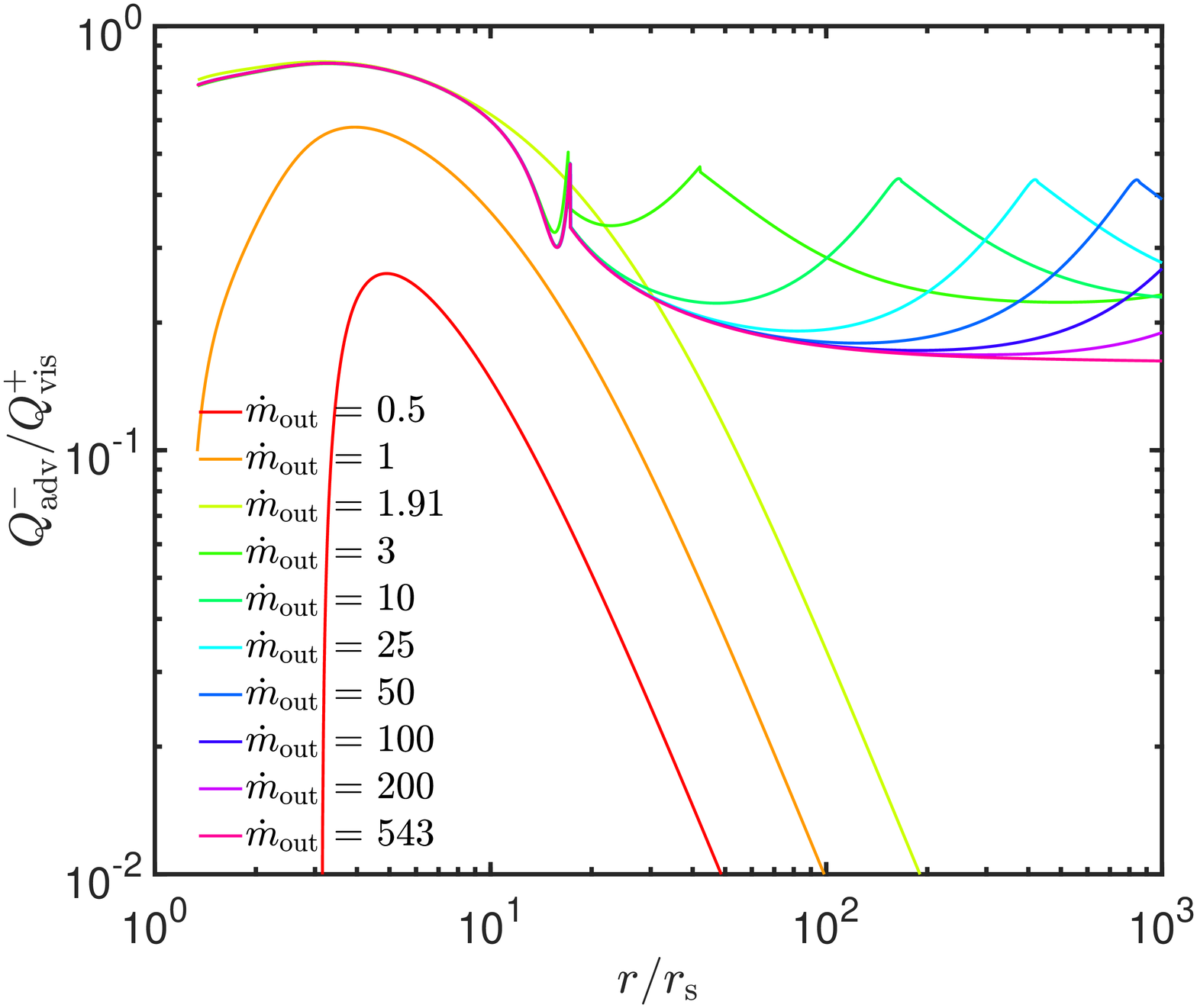}
\caption{The left figure: the same as Figure \ref{fig:m&h10}, but for the radial velocity and sound speed of the disk. The solid lines represent the radial velocity of the gas, while the dashed lines denote the sound speed. The right figure: the ratio of the energy advection to the viscously dissipated energy in the disk.}
\label{fig:vrcs10}
\end{figure*}

\begin{figure*}
	\centering
	\includegraphics[width=8cm,clip=]{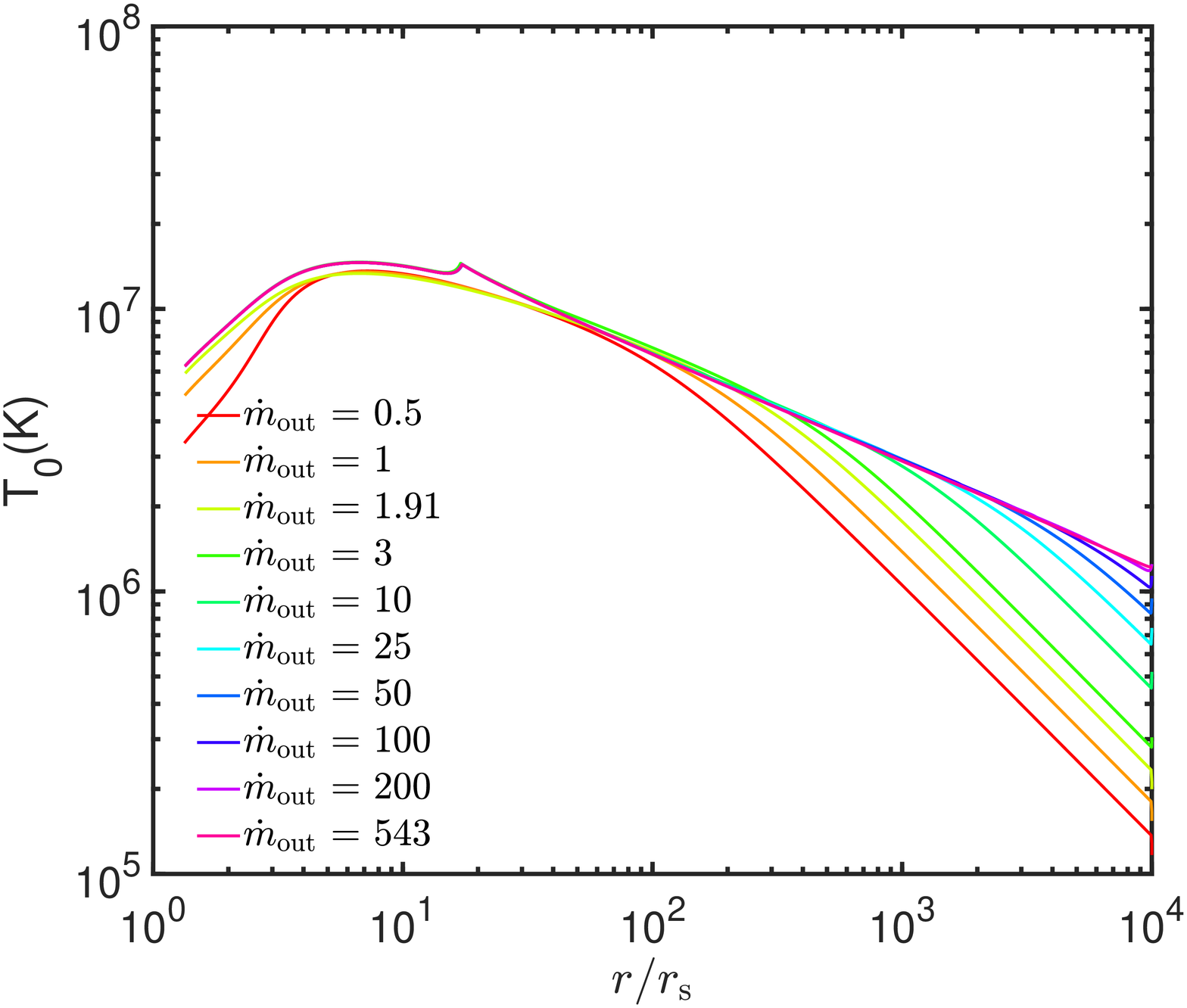}
	\includegraphics[width=8cm,clip=]{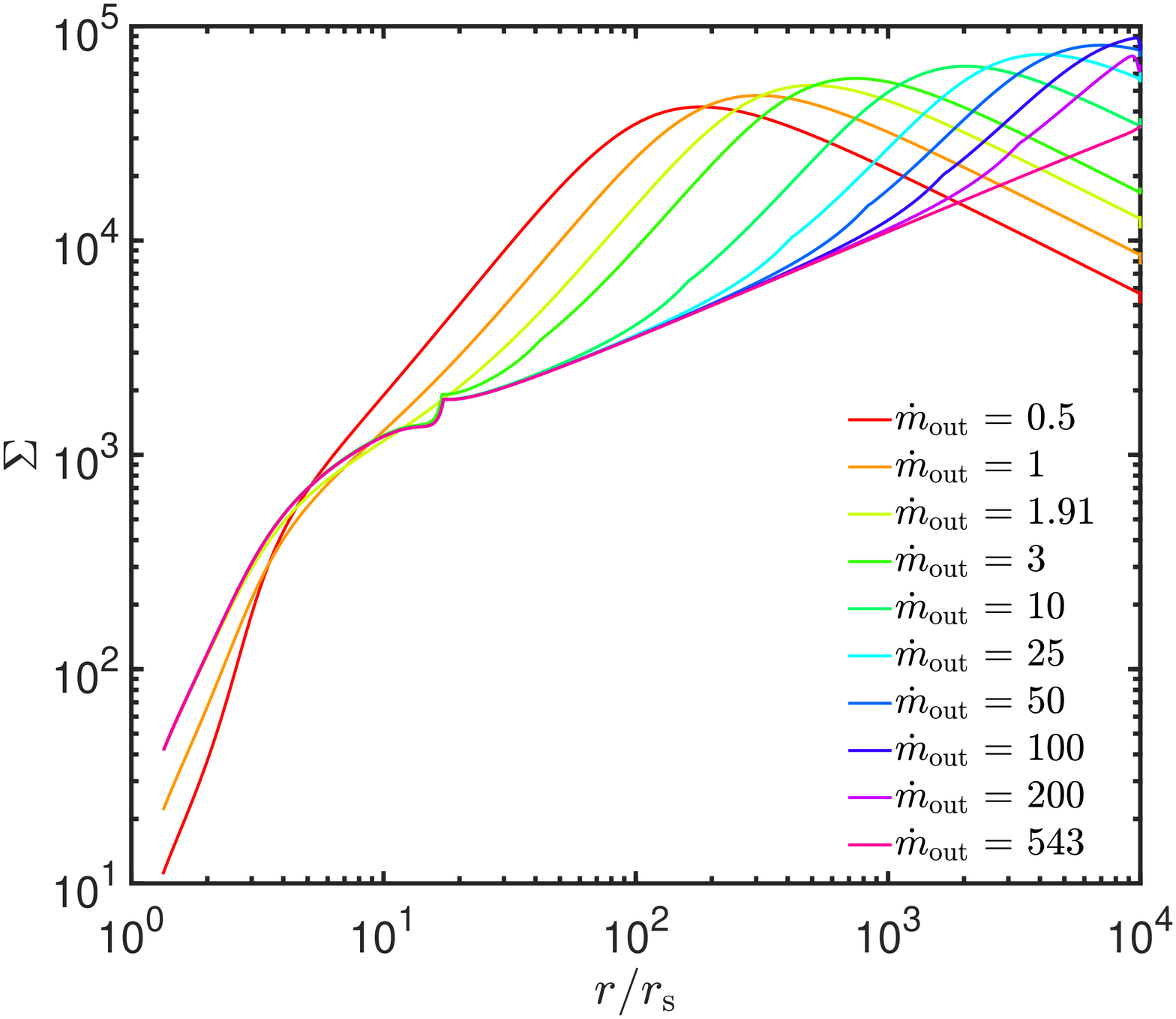}
    \includegraphics[width=8cm,clip=]{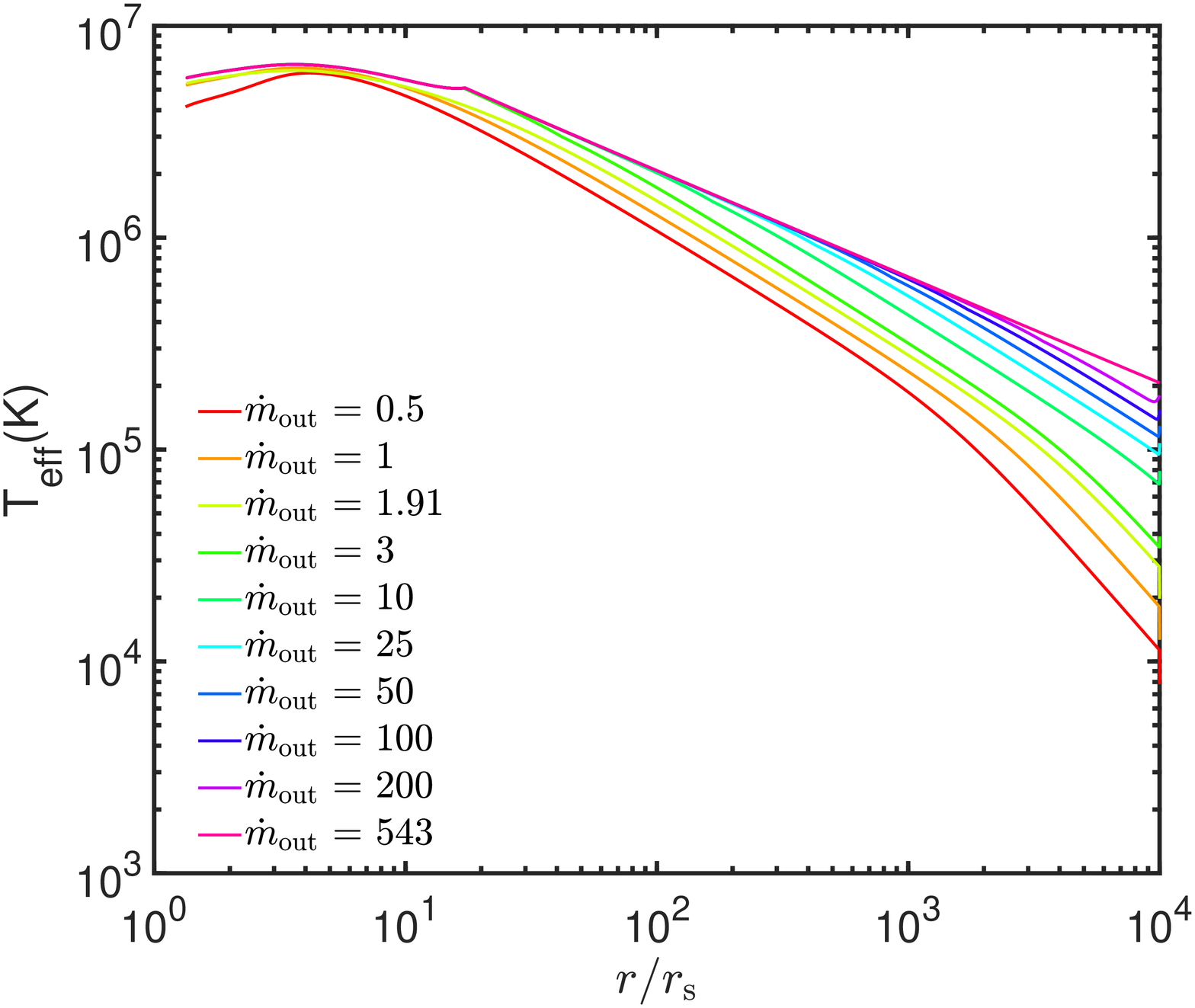}
	\includegraphics[width=8cm,clip=]{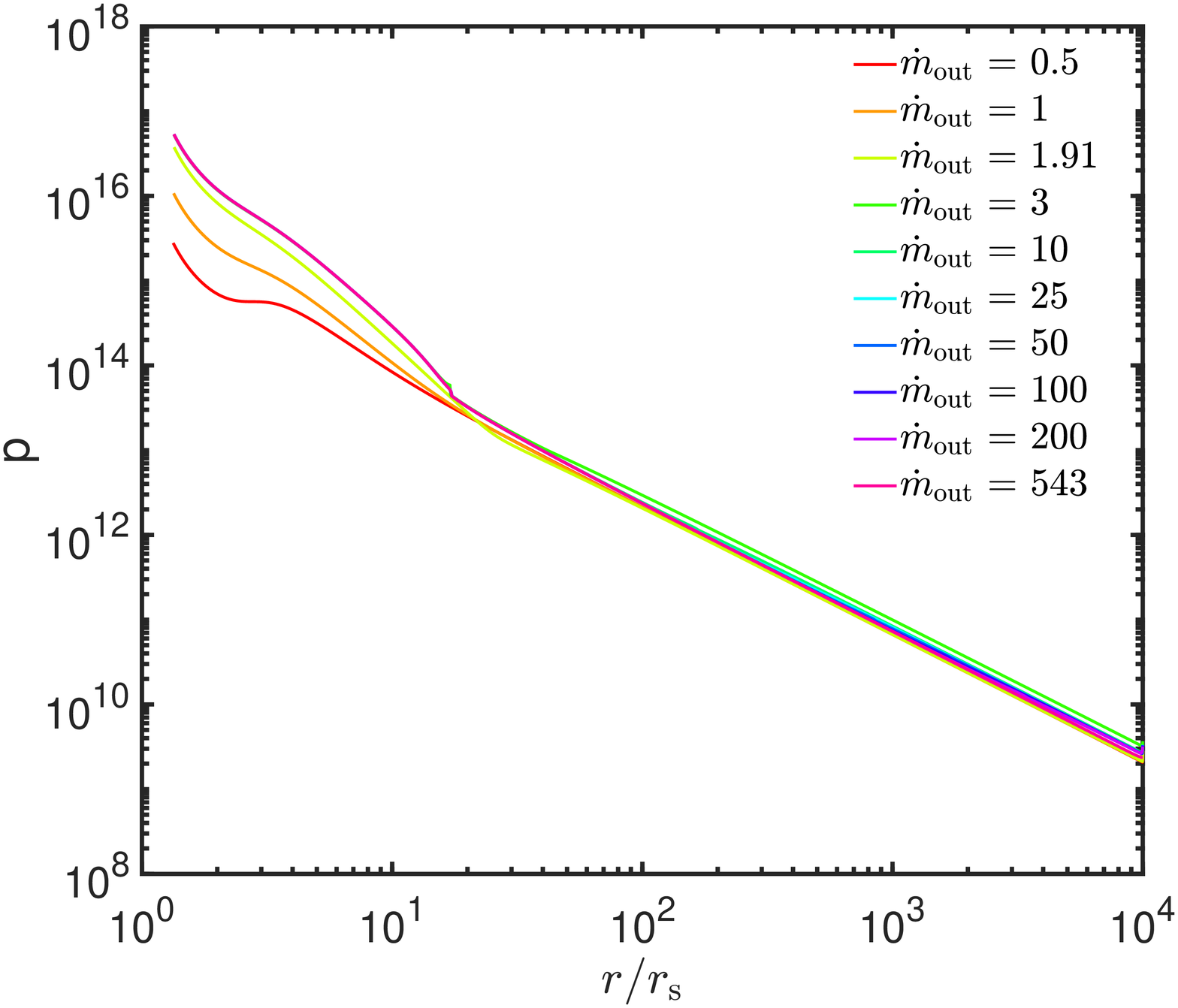}
    \caption{The same as Figure \ref{fig:m&h10}, but for the temperature at the disk mid-plane, the effective temperature, the surface density, and the density of the disk, respectively.}
    \label{fig:state10}
\end{figure*}

\begin{figure}
\centering
\includegraphics[width=8cm,clip=]{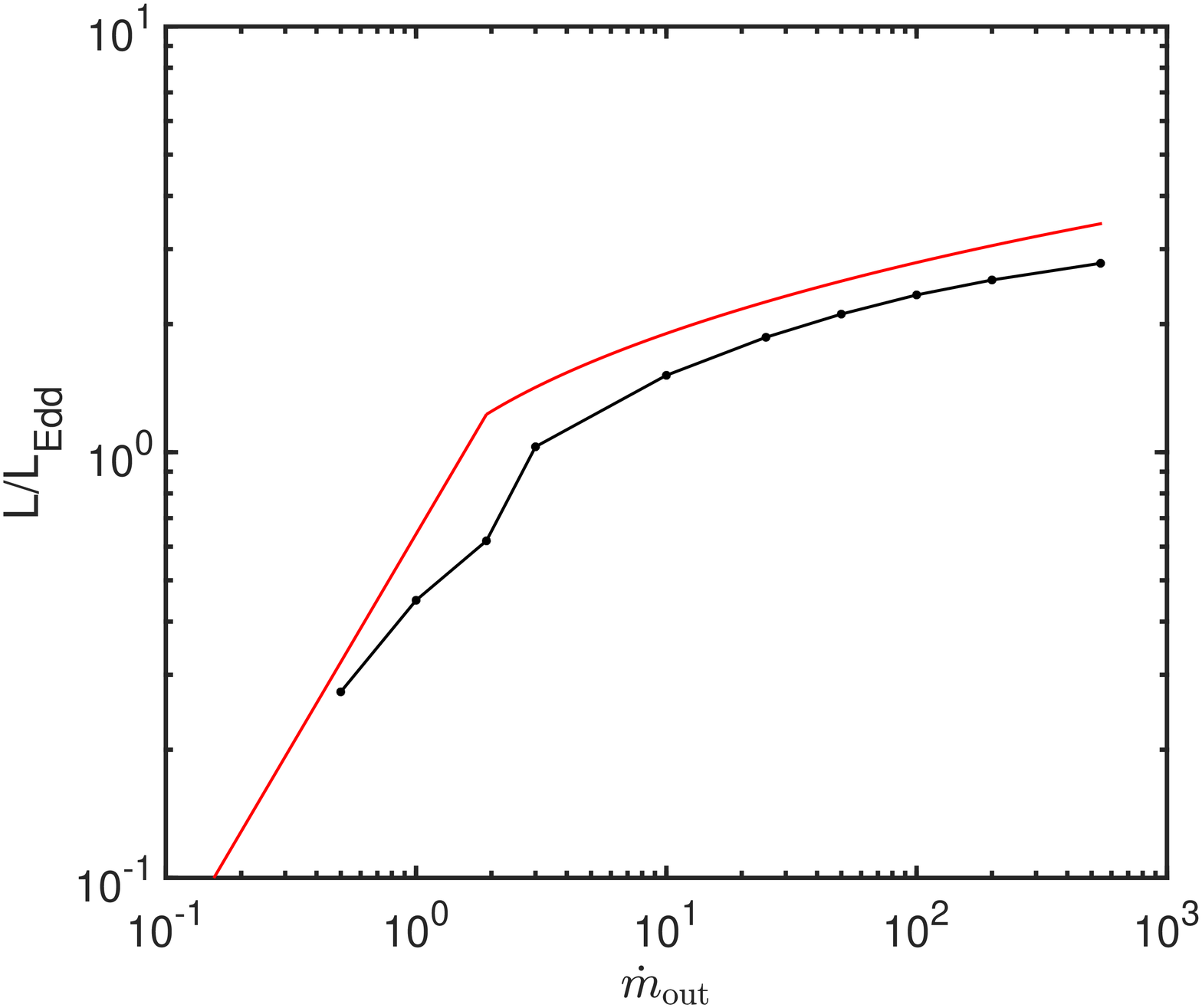}
\caption{The Eddington ratio varies with the mass accretion rate $\dot{m}_{\rm out}$ at $r_{\rm out}$. The black line represents the numerical result of the global structure of the disk, while the red line is the result calculated with the analytical approximation (24) in \citet{2015MNRAS.448.3514C}.}
\label{fig:L10}
\end{figure}

\begin{figure*}

    \centering

    \includegraphics[width=8cm,clip=]{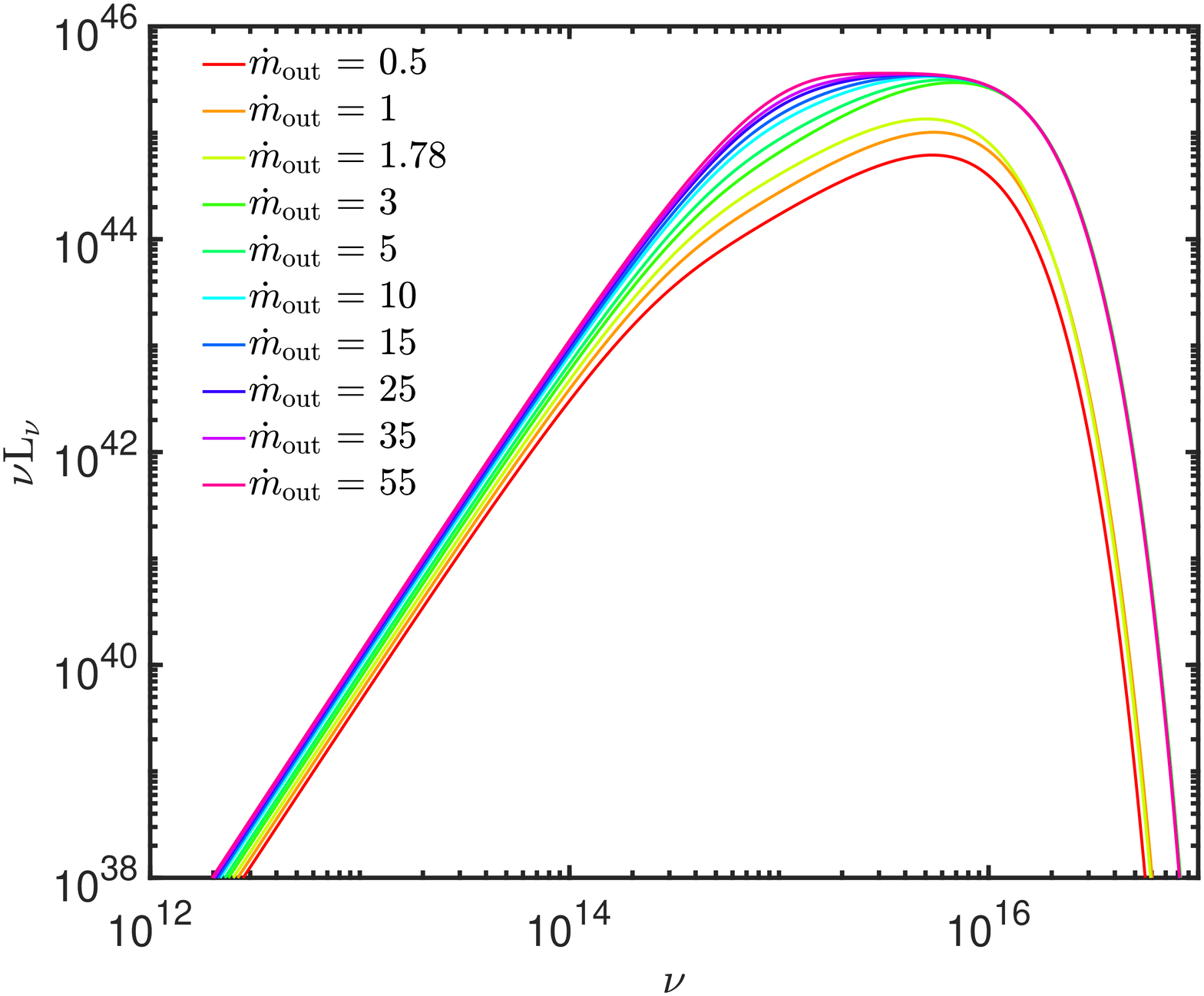}
    \includegraphics[width=8cm,clip=]{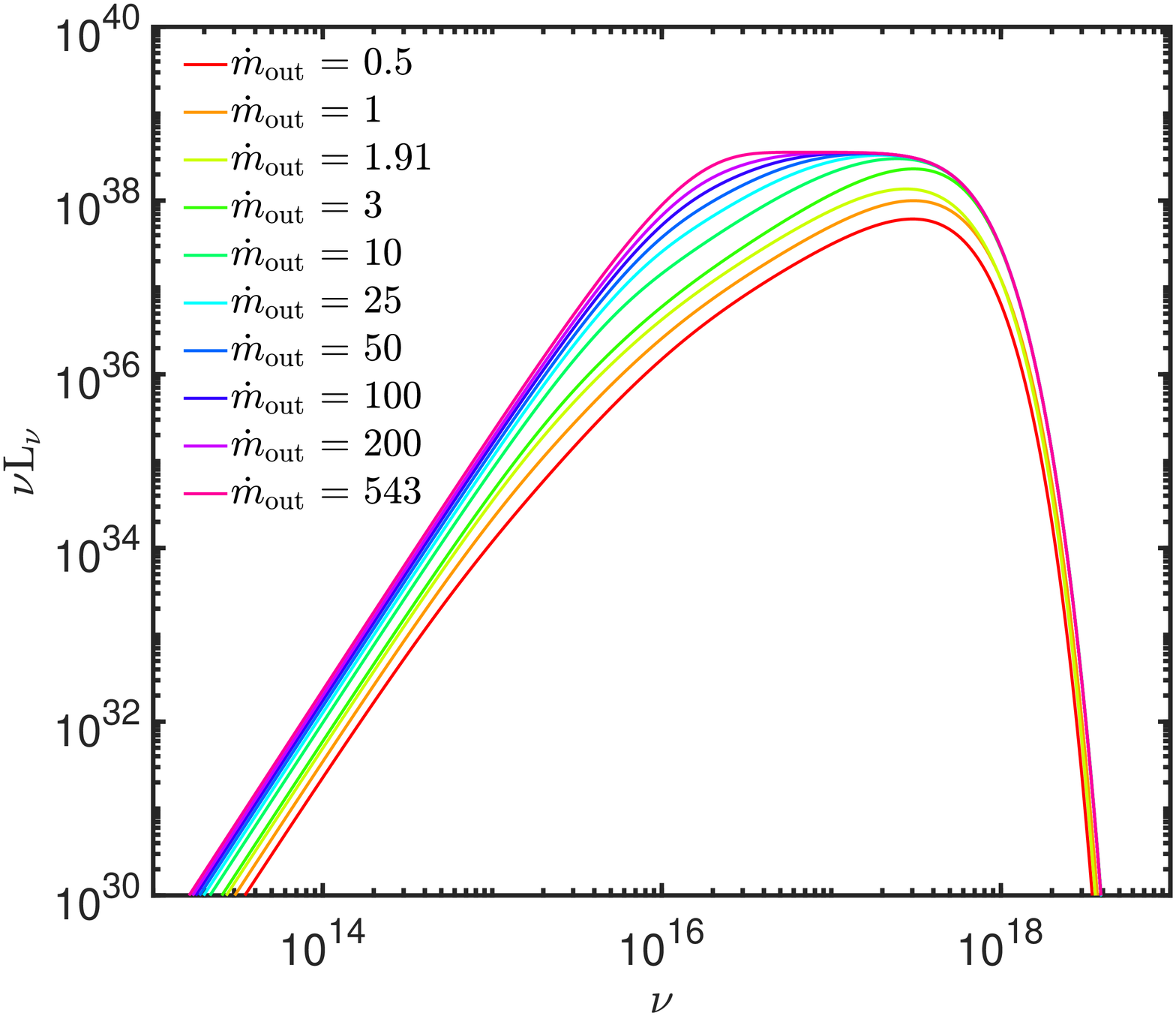}
    \caption{The continuum spectra of the disks with outflows. The left figure is the spectra of the disks surrounding a super-massive black hole with $M=10^8M_\odot$. The right figure is the spectra of the disks surrounding a stellar mass BH with $M=10M_\odot$. }

    \label{fig:spectra}
\end{figure*}

The global solution to a slim disk with outflows derived in this work may also applicable for BH X-ray binaries, so we plot the numerical results for a disk surrounding a stellar mass BH in Figures \ref{fig:m&h10}-\ref{fig:L10}. The dynamical properties of the disk are similar to those of a disk surrounding a massive BH, except its higher temperature compared with its counterpart with a massive BH. This leads to different energy peaks in the continuum spectra (see Figure \ref{fig:spectra}). We find that the continuum spectra of the disk with outflows surrounding a massive BH are saturated in the UV/soft X-ray wavebands, while they are saturated in the hard X-ray waveband for the stellar mass BH cases.




%




\section{Discussion}

In this work, we derive a global solution to a slim accretion disk with radiation driven outflows, in which mass loss rate in outflows and the radial energy advection are properly considered, while the complexity of outflow physics is avoided in the same way as done in \citet{2015MNRAS.448.3514C}. The radiation pressure is in balance with the vertical component of the gravity of the BH in the disk, however, such balance is broken down at the disk surface if the radiation flux is greater than a critical value, and therefore the gas at the disk surface is accelerated into the outflows. This leads to a decrease of the mass accretion rate in the disk, and then the radiation flux decreases. Thus, there are upper limits on the radiation flux and disk thickness, which is regulated by the outflows \citep*[see the detailed discussion in][]{2015MNRAS.448.3514C}.

We find that no outflows are driven from the disk when the mass accretion rate in the disk is lower than the critical value ($\dot{m}<\dot{m}_{\rm crit}$; $\dot{m}_{\rm crit}\simeq 1.78$ for massive BHs, while $\dot{m}_{\rm crit}\simeq 1.91$ for stellar mass BHs). Outflows are driven from a ring region of the disk if the mass accretion rate $\dot{m}_{\rm out}>\dot{m}_{\rm crit}$. The width of the ring increases with the mass accretion rate at the outer edge of the disk. The radiation flux from this ring is limited by the maximal radiation flux $f_{\rm rad}^{\rm max}$, which is a function of radius. The radiation luminosity of the disk with outflows increase with $\dot{m}_{\rm out}$, while the dependence is nonlinear, because part of gas in the disk is driven into outflows and radial energy advection in the disk also plays a role on the disk radiation (see Figures \ref{fig:L1e9} and \ref{fig:L10}).

\citet{2015MNRAS.448.3514C} derived the luminosity of the disk with outflows as a function of the mass accretion rate $\dot{m}_{\rm out}$ at $r_{\rm out}$ without considering the radial energy advection in the disk. It is found that $L/L_{\rm{Edd}}\propto\ln\dot{m}_{\rm out}$, if $\dot{m}$ is large. We find that the numerical results of luminosity of the disk with outflows in this work are slightly lower than the analytic approximation  (Figures \ref{fig:L1e9} and \ref{fig:L10}), which is caused by the advection of the energy in the disk. {It is noted that the energy advection indeed plays some roles in the disk with outflows (the fraction of the energy advection is about $20$ percent for a high mass accretion rate $\dot{m}_{\rm out}$, see Figures \ref{fig:vrcs1e9} and \ref{fig:vrcs10}), which has been neglected in \citet{2015MNRAS.448.3514C}.}

We find that the structure of the inner disk with outflows remains almost same, while the outflow driven region in the disk extends with increasing mass accretion rate $\dot{m}_{\rm out}$. As the effective temperature of the disk is much higher in the inner region of the disk, from which most high energy photons are emitted, so the spectral shape at the high energy end is dominantly determined by the inner region of the disk. It is found that the temperature drops significantly in the inner region of the disk (roughly within the margin stable circular orbit, i.e., $3r_{\rm s}$ for a non-rotating BH), where the gas is plunging onto the BH rapidly. In this region, the dynamic timescale of the gas is much shorter than the radiation timescale, and therefore the advection is dominant over radiation in that region {(see Figures \ref{fig:vrcs1e9} and \ref{fig:vrcs10})}. Only a minor fraction of the gravitational energy released in this region is radiated out, and then the temperature is low, which is similar to the results of a normal slim disk model \citep*[e.g.,][]{1988ApJ...332..646A,2000PASJ...52..133W,2007ApJ...660..541G,2011A&A...527A..17S}.

In this work, we carry out the calculations of global structure of accretion disks with outflows surrounding two different kinds of BHs. It is not surprising that the structure of the disks is quite similar for a massive BH and a stellar mass BH. The effective temperature of the disk surrounding a stellar mass BH is much higher than that for a massive BH, which is a general feature of accretion disks. This leads to the continuum spectra peak at different wavebands (see Figure \ref{fig:spectra}). The effective temperatures
of the disks accreting at high rates almost converge in the inner regions of the disks except the plunging region within $\sim 3r_{\rm s}$ (see Figures \ref{fig:state} and \ref{fig:state10}). Most high energy photons are emitted from the inner region of the disk with $r\ga 3r_{\rm s}$, while the emission from this plunging region contributes little to the continuum spectrum due to its low temperature and small area, it therefore leads to saturation of the continuum spectrum at the high energy end (see Figure \ref{fig:spectra}). We note that the critical mass accretion rate $\dot{m}_{\rm crit}$ is slightly different between a massive BH and a stellar mass BH, i.e., $\dot{m}_{\rm{out}} \sim 1.78$ for a massive BH, and $\dot{m}_{\rm{out}} \sim 1.91$ for a stellar mass BH. We conjecture that such difference arises from the different advection properties (see Figures \ref{fig:L1e9} and \ref{fig:L10}), which may be caused by systematically different temperatures and densities of the disks for these two kinds of BHs. The saturation of the emission in the high-energy end of the disk spectra can be tested by the observations of luminous objects. It was even suggested that the saturation of the spectra of luminous quasars can be used as a type of `cosmological candles' with no need to search massive
black holes accreting at extremely high rates \citep*[e.g.,][]{2013PhRvL.110h1301W}.

Supermassive black holes have been discovered at high redshifts, some of which even with $z \gtrsim 7$, at the age of the Universe less than 0.6 Gyr \citep{2011Natur.474..616M,2015Natur.518..512W,2016ApJ...833..222J,2018Natur.553..473B,2018ApJ...869L...9W,2018arXiv181111915Y,2019ApJ...872L...2M}.  It is believed that the black holes growing to $ \sim 10^9 \rm{M}_{\odot}$ in such a short period needs both large seed black holes
($\gtrsim 10^3 \rm{M}_{\odot}$) and extremely high accretion rates. However, our calculations show that luminous disks accreting at high rates inevitably drive outflows, which set an upper limit on the rate of mass accreted by the massive BH. The BH mass growth rate is limited by $\dot{m} \la 2$. Our results seem to support the scenario that the BH growth through accreting persistently at a rate close to the Eddington value \citep*[e.g.,][]{2005ApJ...620...59S,2009ApJ...696.1798T,2014CQGra..31x4005T}.

In this work, we focus on the global solution to a slim disk with outflows, and the detailed dynamics of the outflows have not been considered, which is beyond the scope of the present work. The continuum spectra of the disks are calculated based on multi-color black body assumption. We believe the present calculations can indeed describe the main features of the continuum spectra, such as, saturation in the high-energy end of the spectra. More detailed calculations based on the derived global structure of the disks with outflows by including the detailed radiative transfer in the disks with electron scattering effect could be carried out in the future in order to confront with the spectral observations of luminous objects, which are beyond the scope of this work.



\acknowledgments
\section*{Acknowledgments}

We thank the referee for his/her helpful comments. This work is supported by the NSFC (grants 11773050, 11833007 and 11573023), and the CAS grant (QYZDJ-SSWSYS023).

\end{document}